\documentclass[acmsmall, natbib=false, screen,authordraft,nonacm,review=false,timestamp=false]{acmart}
\usepackage{preprint_header}
\usepackage[backend=biber, style=numeric, sorting=none]{biblatex} 
\usepackage{multirow}
\usepackage{tabularx}
\usepackage{longtable}
\AtBeginDocument{%
 \providecommand\BibTeX{{%
  \normalfont B\kern-0.5em{\scshape i\kern-0.25em b}\kern-0.8em\TeX}}}

\AtBeginDocument{%
\thispagestyle{preprintbox}

  \providecommand\BibTeX{{%
    Bib\TeX}}}

\acmDOI{10.1145/3677092}

\acmConference[CHI PLAY '24]{ACM Hum.-Comput. Interact., Vol. 8, No. CHI PLAY, Article 327}{October 14--17, 2024}{Tampere, Finland}
\acmJournal{PACMHCI}
\acmVolume{8}
\acmNumber{CHI PLAY}
\acmArticle{327 (Conditionally accepted)}
\acmYear{2024}
\acmMonth{10}

\begin{document}
\thispagestyle{preprintbox}
\title{News Ninja: Gamified Annotation of Linguistic Bias in Online News}
\thispagestyle{preprintbox}

\author{Smi Hinterreiter}
\orcid{0000-0002-7029-2753}
\affiliation{%
\institution{University of Würzburg}
 \city{Würzburg}
 \country{Germany}
}
\author{Timo Spinde}
\orcid{0000-0003-3471-4127}
\affiliation{%
 \institution{University of Göttingen}
 \city{Göttingen}
 \country{Germany}
}
\author{Sebastian Oberdörfer}
\orcid{0000-0002-8614-1888}
\affiliation{%
 \institution{University of Würzburg}
 \city{Würzburg}
 \country{Germany}
}
\author{Isao Echizen}
\orcid{0000-0003-4908-1860}
\affiliation{%
 \institution{National Institute of Informatics}
 \city{Tokyo}
 \country{Japan}
}
\author{Marc Erich Latoschik}
\orcid{0000-0002-9340-9600}
\affiliation{%
 \institution{University of Würzburg}
 \city{Würzburg}
 \country{Germany}
}

\renewcommand{\shortauthors}{Hinterreiter, Spinde, Oberdörfer, Echizen, and Latoschik}

\begin{abstract}
Recent research shows that visualizing linguistic bias mitigates its negative effects.
However, reliable automatic detection methods to generate such visualizations require costly, knowledge-intensive training data.
To facilitate data collection for media bias datasets, we present News Ninja, a game employing data-collecting game mechanics to generate a crowdsourced dataset.
Before annotating sentences, players are educated on media bias via a tutorial.
Our findings show that datasets gathered with crowdsourced workers trained on News Ninja can reach significantly higher inter-annotator agreements than expert and crowdsourced datasets with similar data quality.
As News Ninja encourages continuous play, it allows datasets to adapt to the reception and contextualization of news over time, presenting a promising strategy to reduce data collection expenses, educate players, and promote long-term bias mitigation.
\end{abstract}



\received{21 February 2024}
\received[revised]{03 June 2024}
\received[accepted]{5 July 2024}

\begin{teaserfigure}
 \includegraphics[width=\textwidth]{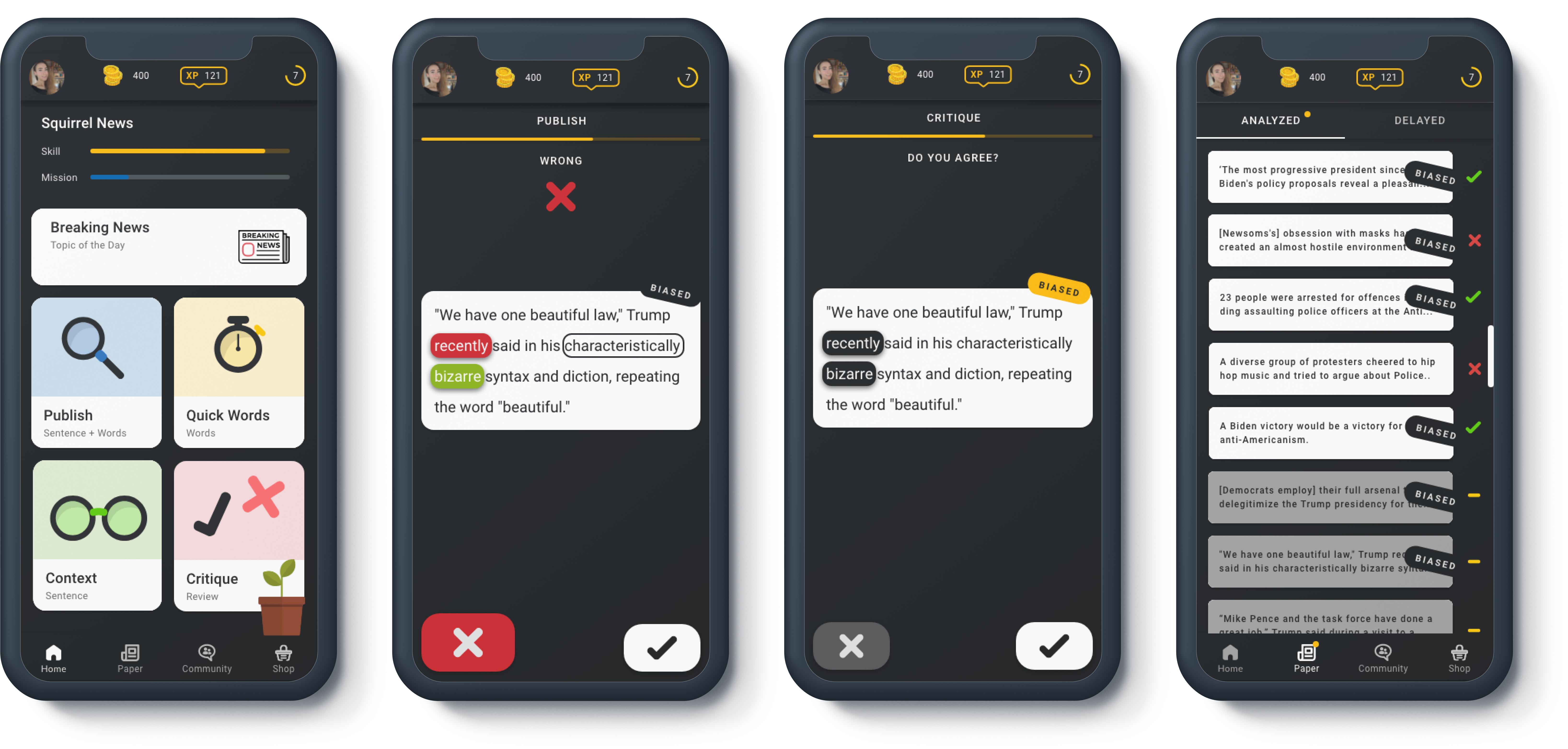}
 \caption{From left to right, the first screen shows News Ninja's home screen with player metrics, the group mission, the breaking news tile, and four game modes. The second screen shows feedback on an incorrectly annotated sentence in the \textit{Publish} game mode (Section \ref{sec:gamemodes:details}), with one correct word, one incorrect word, and one missed word. The third screen displays the \textit{Critique} mode with the same annotated sentence. Players can agree or disagree by swiping or using the buttons. The last screen shows the \textit{Paper} section where played sentences are archived. Sentences missing a ground truth during play show up white when players can collect feedback.}
 \Description{From left to right, the first screen shows News Ninja's home screen with the profile picture and the player metrics money, XP, and level in the top bar. All screens display the top bar and a black background. The main interface shows the skill level bar, the group mission, the breaking news tile, and the four game modes: Publish, Quick Words, Context, and Critique. Below is a bottom navigation bar with the home button on the left. It is followed by the Paper button, Community button, and Shop button. The second screen shows feedback on an incorrectly annotated sentence card with the sentence "We have on beautiful law, Trump recently said in his characteristically bizarre syntax and diction, repeating the word beautiful." "Bizarre" is highlighted green to indicate a correct annotation, "recently" is highlighted red to indicate an incorrect annotation, and "characteristically" has a black outline to indicate a missed word. A red cross on top indicates a miss of sentence level bias. On the bottom of the screen is a red cross button the player just pressed on the left and a white checkmark button on the right. The first leads players to the discussion of the sentence. The second lets them collect their reward. The third screen displays the Critique mode with the same annotated sentence. Above the sentence card, a text reads, "Do you agree?" Players can agree or disagree with a yellow bias label attached to the sentence card and the black highlighted biased words by swiping the card or using the buttons below (a cross on the left and a checkmark on the right). A progress bar below the top bar shows their progress in the ten sentences. The last screen shows the Paper section where played sentences are listed below each other. Sentences that have not had a ground truth during play show up white when an agreement is reached, and players can collect their feedback. Biased sentences obtain a black bias label on their right side. Next to it, on the outer right of the screen, are symbols indicating hits (green checkmark), misses (red cross), or a waiting-for-feedback status (yellow dash, delayed feedback).}
 \label{fig:teaser}
\end{teaserfigure}

\maketitle
\thispagestyle{preprintbox}

\section{Introduction}
Online news is the predominant information source for current events and critical political issues \cite{dallmann2015a, eveland2003impact, norrisVirtuousCirclePolitical2000} and is generally perceived as trustworthy \cite{spindeAutomatedIdentificationBias2021}.
However, news media frequently carry inherent biases, a structural defect \cite{mediaBiasDefinitionDomke1999} that can influence public opinion \cite{effectsArd2017, mediaBiasVotingDellaVigna2007, oneBiasFitsAllEberl2017}.
As media bias is a multifaceted concept with various subtypes \cite{spindeIntroducingMediaBias2022}, we focus on linguistic bias.
Linguistic bias is evident in word choice and framing \cite{fairBudak206, PhrasingBiasHube2019, mullainathanSpinMediaBias2002} and describes the usage of language to convey a certain view of events, groups, or individuals \cite{spindeIntroducingMediaBias2022}.
Readers are often oblivious to this bias, which can lead to a compromised understanding of issues and promote a skewed perspective \cite{spindeEnablingNewsConsumers2020, ribeiroMediaBiasMonitor2018, FramingClimateUncertaintyGaissmaier2019}.
One effective countermeasure against the adverse effects of linguistic bias is highlighting its presence to readers \cite{spindeEnablingNewsConsumers2020, spindeHowWeRaise2022}.
While possible, manual annotation of bias by experts is impractical given the sheer volume of news content \cite{spindeIntegratedApproachDetect2020}.
Recent advancements in natural language processing (NLP) offer promise for automated bias detection \cite{lei-etal-2022-sentence, spindeNeuralMediaBias2021, Wessel2023}.
Yet, their current performance still falls short of the required performance for end-user solutions \cite{spindeNeuralMediaBias2021}.
This primarily arises from the resource-intensiveness and complexity of creating large, high-quality training datasets \cite{Hamborg2019inter}.
Biases such as spin or framing bias complicate the detection task due to their reliance on context \cite{spindeIntroducingMediaBias2022}.
For example, the word "killed" is unbiased in the sentence "12 people were killed."
However, it becomes biased in the sentence "Trump killed his opponent in his speech."

Games With a Purpose (GWAPs) generate data and labels through human computation as a byproduct of gaming \cite{vonahnGamesPurpose2006, madgeIncrementalGameMechanics2019, poesioPhraseDetectivesUtilizing2013}.
Concurrently, games can serve an educational function and increase awareness of the game's topic \cite{susiSeriousGamesOverview2007, basolGoodNewsBad2020}.
Merging those concepts could increase player comprehension of the issue, thereby facilitating the collection of higher-quality data \cite{madgeIncrementalGameMechanics2019}.
Thus, GWAPs present a viable solution to the three challenges of increasing players' bias detection skills, making annotation tasks engaging, and circumventing the need for experts for dataset creation.
Specifically for a linguistic bias GWAP that requires the education of players to annotate biased words and sentences accurately, we investigate the research questions:
\begin{itemize}
    \item (Q1) How can knowledge of linguistic bias be transferred in an interactive and gamified manner?
    \item (Q2) How can game mechanics facilitate the annotation task?
    \item (Q3) Can player-generated data achieve comparable results to expert-generated data?
\end{itemize}

This work introduces the GWAP News Ninja (Figure \ref{fig:teaser}).
News Ninja aims to increase players' linguistic bias detection skills before gathering data to refine automated bias detection.
The design process addresses Q1 by employing game design frameworks to translate annotation guidelines \cite{spindeNeuralMediaBias2021} into game elements, including a storyline, progression elements, direct feedback, rewards, and social elements (Figure \ref{fig:derivatives}) 
\cite{kirschnerTenStepsComplex2017, oberdorferGamifiedKnowledgeEncoding2018,chouActionableGamificationPoints2015, arnabMappingLearningGame2015, segundodiazBuildingBlocksCreating2022}.
For (Q2), we deconstruct the annotation process into game mechanics, which we qualitatively pre-test in combination with the tutorial (Section \ref{sec:pretest}).
Based on the results, we re-design the tutorial and label-generating game modes (Section \ref{sec:gamemodes}).
To assess data quality for (Q3), we develop and test News Ninja as a streamlined, responsive web application to compare player-generated data to expert-generated data (Section \ref{sec:studydesign:outline}).

Player-generated labels show a significant 10.28\% increase in inter-annotator agreement (IAA) compared to the baseline dataset BABE \cite{spindeNeuralMediaBias2021}.
Experts\footnote{Experts are three researchers with more than one year of experience in media bias research from the university network.} re-annotate the game sentences and achieve an agreement of 79.8\%, indicating News Ninja as a promising approach for crowdsourcing expert-like linguistic bias labels.
The prototype showcases the possibilities of combining game design, education, and annotation tasks.
As the first functional GWAP focused on linguistic bias, News Ninja introduces a GUI, tutorial, five game modes, a feedback system, and a delayed feedback mechanism for sustained player engagement when a ground truth is missing.
The game can be adapted to various NLP annotation scenarios to create various linguistic datasets.

How news is perceived and contextualized is subject to continuous evolution, yet datasets remain static.
Applications like News Ninja hold the potential to update datasets to mirror these changes while simultaneously enhancing public awareness and bias detection skills.
We conclude by discussing our game annotation mechanics for subjective truths and potential cognitive biases within the dataset.
The dataset is publicly available.\footnote{\url{https://github.com/Media-Bias-Group/News-Ninja}}

\section{Background} \label{sec:bg}
Linguistic bias is reflected in statements when language is systematically used to reinforce stereotypes or specific perceptions of groups, events, or individuals \cite{Beukeboom_2017}.
This form of bias is fine-grained, evident in individual words, phrases, or sentence structures, and can alter the context and meaning of a statement.
Such bias can manifest through one-sided terms or adjectives that amplify or add subjective meaning to a sentence or text \cite{Recasens_2013}.
The choice of words can influence the perceived credibility of a statement \cite{Recasens_2013}.
Additionally, words and phrases often carry connotations, integrating subtle feelings or biases into statements \cite{Rashkin_2016}.

\subsection{Automatic Detection of Linguistic Bias} \label{sec:bg:detection}
NLP classifiers show potential in algorithmically detecting and indicating linguistic bias \cite{lei-etal-2022-sentence, spindeNeuralMediaBias2021, Wessel2023}.
The currently most common approach is fine-tuning large language models with bias datasets containing statements or sentences \cite{spindeIntroducingMediaBias2022}.
However, their accuracy falls short of the standards required for consumer tools \cite{spindeIntroducingMediaBias2022} because extensive, high-quality bias datasets are missing \cite{spindeMBICMediaBias2021, spindeIntroducingMediaBias2022}.
Due to the complex nature of the annotation task, crowdsourced datasets exhibit low agreement and higher noise \cite{spindeMBICMediaBias2021}.
\textcite{spindeMBICMediaBias2021}'s crowdsourced dataset MBIC achieves a F1-score of 0.43 and an IAA of $\alpha$ = 0.21.
Contrasting, expert datasets are costly but achieve a higher F1-score of 0.804 and an IAA of $\alpha$ = 0.39 \cite{spindeNeuralMediaBias2021}.
Prior research tries to optimize cost and reliability by balancing crowdsourced and expert labels and training non-experts over extended periods to become experts \cite{spindeNeuralMediaBias2021}.
However, this method remains costly for large-scale application \cite{spindeNeuralMediaBias2021}, stressing the need for alternative solutions in generating reliable media bias datasets.
Our strategy addresses the challenges by educating non-experts and substituting financial incentives with engaging gameplay.

\subsection{Media Literacy Education and Online Civic Reasoning Approaches} \label{sec:bg:literacy}
We investigate interactive education methods in bias and media literacy to design game mechanics that instruct and train players \cite{basolGoodNewsBad2020, axelssonLearningHowSeparate2021}.
Educational research on media bias itself is sparse and merely focuses on the impact and the perception of media bias \cite{cook_neutralizing_2017, spindeEnablingNewsConsumers2020, spindeHowWeRaise2022}.
Curricula on media literacy \cite{aufderheideMediaLiteracyReport1993a} and online civic reasoning (OCR) \cite{mcgrewLearningEvaluateIntervention2020, axelssonLearningHowSeparate2021} touch upon bias through agenda-setting --- which news ultimately gets reported --- and framing --- how and with which words and phrases the news is presented \cite{buckinghamTeachingMediaPosttruth2019}.
Generally, interventions related to News Ninja target younger demographics through school curricula \cite{mcgrewLearningEvaluateIntervention2020} or employ interactive checklists \cite{GlassmanWHO2022}, with interactive online courses \cite{axelssonLearningHowSeparate2021} and serious games \cite{basolGoodNewsBad2020}.
For instance, the interactive web-based learning tool "The News Evaluator" equips users with skills for critical online content engagement \cite{axelssonLearningHowSeparate2021}.
The application packs OCR objectives into a structured format, a tutorial, hands-on learning tasks, and implicit and explicit feedback.

"Bad News" is a serious game for media literacy education \cite{basolGoodNewsBad2020}.
It teaches six misinformation strategies by embedding them into the story and letting players adopt the antagonist's role.
Players build their media empire's following and credibility by generating and disseminating news, enhanced by a gameful interface, ownership elements, and achievements to foster motivation and learning.
The game significantly improves players' ability to detect misinformation through inoculation by exposing them to weak doses of misinformation to build cognitive immunity \cite{cook_neutralizing_2017, basolGoodNewsBad2020}.

\subsection{Games for Data Collection} \label{sec:bg:games}
Integrating insights from Games With a Purpose (GWAPs), News Ninja's design objectives aim to educate while simultaneously crowdsourcing linguistic bias labels, introducing a  unexplored strategy in media bias research \cite{madgeIncrementalGameMechanics2019, hinterreiterGamifiedApproachAutomatically2021}.
Von \textcite{vonahnGamesPurpose2006} describes GWAPs as games that use human computation \cite{Law2011} to produce data during gameplay \cite{vonahnGamesPurpose2006, vonahnDesigningGamesPurpose2008, Ahn2004LabelingImages, law2007tagatune, Ahn2009InputAgreement}.
\textcite{SeriousGamesImproved} view it as crowdsourcing via games.
They offer inexpensive and scalable data collection \cite{vonahnGamesPurpose2006} with an inclusive design that appeals to casual gamers \cite{madgeIncrementalGameMechanics2019}.
Broadly, GWAPs are serious games, defined as games with an objective beyond pure entertainment.
They often intersect with education or simulation \cite{susiSeriousGamesOverview2007} and can enhance educational scenarios \cite{GroveSeriousGames2010, VanEckGBL2006, GeeGamesLearningLiteracy2003, GreitzerCognitiveScienceForGamification2007}.
In contrast to gamification, which applies game elements in non-gaming contexts \cite{DeterdingGamificationSeriousGames2011, krathRevealingTheoreticalBasis2021}, serious games are comprehensive gaming experiences.
Subsequently, GWAPs can fall anywhere on the spectrum between serious games and gamification \cite{tuiteGWAPsGamesProblem2014}.
Thus, sustaining player engagement remains a challenge; GWAPs must be compelling, and players must be able to complete the task \cite{madgeTestingTileAttackThree2018}.
This stresses the need for engagement strategies to ensure sustainable, long-term data collection to account for changes in news content \cite{madgeIncrementalGameMechanics2019}.

Historically, GWAPs that collect training data have been used for tasks associated with more objective truths, such as image labeling \cite{raddickGalaxyZooExploring2010, vonahnGamesPurpose2006} or grammatical parts-of-speech tagging \cite{madgeIncrementalGameMechanics2019, kicikogluWormingoTrueGamification2019}.
GWAPs that focus on subjective, cultural truths, such as detecting abusive, stereotypical, or sexist language, often merely describe their systems \cite{bonetti3DRolePlayingGame2020, millourKatanaGrandGuru2019} or conduct UX studies \cite{althaniLessTextMore2022, Bry2020NewsroomAG}.
Few evaluations manually assess the data \cite{aghelmalekiGeneratingNonsexistCorpus2019, Bry2020NewsroomAG}, use metrics like inter-annotator agreement (IAA), or compare with a domain-specific gold standard \cite{vannellaValidatingExtendingSemantic2014, levitanLieCatcherGameFramework2020}, hindering direct comparison.
Therefore, viable ways to evaluate a linguistic bias GWAP include assessing UX, manually evaluating the data, and comparing GWAP labels to gold standard datasets.

\subsection{Combining Game Mechanics for Learning and Data Collection} \label{sec:bg:takeaways}
Our goal is to merge learning with data collection within a single game, aiming for players to develop mental models for bias detection applicable across various tasks and in real-world scenarios \cite{oberdorferGamifiedKnowledgeEncoding2018}.
We employ the "Gamified Knowledge Encoding Model" to leverage interacting game mechanics to help players internalize learning objectives \cite{oberdorferGamifiedKnowledgeEncoding2018}.
Those game mechanics transform learning objectives into game elements while preserving the essence of fun and engagement \cite{arnabMappingLearningGame2015}.
Game mechanics serve as the interface between player interactions and learning outcomes, enclosing both game-bound elements tied to the storyline and game principles and player-bound actions executed by players.
Through these interactions, players can acquire declarative knowledge and, with sufficient repetition, procedural knowledge \cite{oberdorferGamifiedKnowledgeEncoding2018}.

To implement learning and data collection mechanisms, we analyze two games that already combine them \cite{madgeMakingTextAnnotation2019, vonahnDuolingoLearnLanguage2013}.
The language learning application Duolingo, originally designed to train users to translate language segments, faces similar challenges of training players to become experts \cite{spindeNeuralMediaBias2021} and sustaining their engagement for continuous translation \cite{vonahnDuolingoLearnLanguage2013}.\footnote{Initially, Duolingo's primary objective was translation. However, it pivoted to language learning with a subscription-based business model.}
Duolingo resolves the conflict inherent in GWAPs - balancing data collection with providing challenging, educational, and entertaining experiences to players - by optimizing motivation through learning opportunities, gamification, challenges, and social nudges \cite{madgeIncrementalGameMechanics2019}.
They further incorporate the pedagogical agent Duo \cite{baylorPedagogicalAgentDesign2004}, who guides players through the lessons.
Pedagogical agents are essentially characters in a virtual learning landscape.
They serve diverse instructional roles by providing help, guidance, and assistance in learning \cite{marthaDesignImpactPedagogical2019} through social cues that can cause social responses \cite{atkinsonFosteringSocialAgency2005}.

Similarly, the GWAP WordClicker educates players about word classes and gathers word class labels \cite{madgeMakingTextAnnotation2019}.
Players collect words of a chosen class to fill up jars used as resources for in-game currency production.
They progressively learn to identify different classes by accumulating currency and buying new word classes.
A time component increases the challenge for players.
WordClicker serves as a tutorial for the more complex GWAP Tile Attack \cite{madgeTestingTileAttackThree2018} and is part of \textcite{madgeLingoTownsVirtualWorld2022a}'s pipeline of games for part-of-speech tagging.
Each game in the sequence augments complexity and cumulatively enhances player progression and engagement.

An effective linguistic bias GWAP integrates a structured, hands-on, expert-reviewed tutorial, complemented by learning mechanics with a cohesive feedback system and testing tasks for player qualification \cite{hinterreiterGamifiedApproachAutomatically2021}.
It involves setting explicit goals and weaving a compelling narrative, ideally situated within a context related to the learning objective \cite{oberdorferBetterLearningGaming2021}.
The feedback enables iterative learning from errors and fosters improvement through repetition and correction, while testing tasks facilitate data selection for the final dataset.
The learning material and tasks should be split into understandable units using didactic content structuring.
For linguistic bias, we follow \textcite{spindeNeuralMediaBias2021} by taking sentences from news articles and collecting annotations at both the sentence and word levels.
Players annotate sentences as "biased" or "not biased," while they can mark individual words as "biased" (Figure \ref{fig:GameMechanic}).
To enhance player enjoyment, we incorporate game elements outlined for serious games and GWAPs by \textcite{segundodiazBuildingBlocksCreating2022}, including feedback, progression elements such as levels, rewards like currency, and social interactions through discussion threads, as detailed in Section \ref{sec:motivational}.
The combination of a captivating interface and a progressively challenging GWAP, stressing the underlying purpose, can heighten player motivation and ensure their sustained engagement \cite{kosterTheoryFunGame2013, oberdorferBetterLearningGaming2021, chouActionableGamificationPoints2015, tuiteGWAPsGamesProblem2014}.

Notably, the typical player base of GWAPs belongs to the \textit{Achiever} or \textit{Philanthropist} player group \cite{tondelloGamificationUserTypes2016}.
Implementing game mechanics that spotlight performance, milestones, progression, or the overarching mission drives their motivation \cite{tondelloElementsGamefulDesign2017, chouActionableGamificationPoints2015, segundodiazBuildingBlocksCreating2022}.
Such mechanics can promote prolonged player engagement \cite{chouActionableGamificationPoints2015} by fostering enjoyable experiences and flow states \cite{csikszentmihalyiFlowPsychologyOptimal1990}.
Suited mechanics make the action steps harder, show progress and achievements, or unlock new content and interaction possibilities.
While it is essential to design a linguistic bias GWAP with broad appeal to diversify the dataset, strategically fostering the motivations of \textit{Achievers} and \textit{Philanthropists} can amplify their contributions \cite{tuiteGWAPsGamesProblem2014}.
This becomes evident when considering that 3\% of players are responsible for producing between 80-90\% of the data \cite{madgeMakingTextAnnotation2019}.
Players' backgrounds impact bias perception, so GWAPs must include them to ensure a balanced, diverse, and minimally biased dataset \cite{drawsChecklistCombatCognitive2021}.
The mission statement of the game, which explains the deeper purpose behind the annotation task and stresses the societal importance, can include the reasoning for querying player demographics.
Stressing the learning value for players themselves can further increase intrinsic motivation.

\section{The News Ninja Game} \label{sec:game}
News Ninja is a GWAP designed to educate players and collect linguistic bias annotations on words and sentences.
The game translates written annotation guidelines into an interactive tutorial (Q1) and converts the annotation process into two data game mechanics (Q2).
News Ninja's system design builds from four primary components: (1) Two data annotation mechanics (Section \ref{sec:mechanic}), (2) general game mechanics\footnote{All game mechanics are detailed in Table \ref{tab:allmechanics} and explained in Section \ref{sec:motivational}.},
and their integration within (3) the tutorial (Section \ref{sec:tutorial}) and (4) five game modes (Section \ref{sec:gamemodes}).
The two data mechanics extract annotations from player interactions to aggregate them into bias labels.
The tutorial aims to increase players' bias detection skills, introducing the data mechanics to prepare players for the five game modes.

\begin{figure}[!htbp] \centering
 \includegraphics[width=0.78\textwidth]{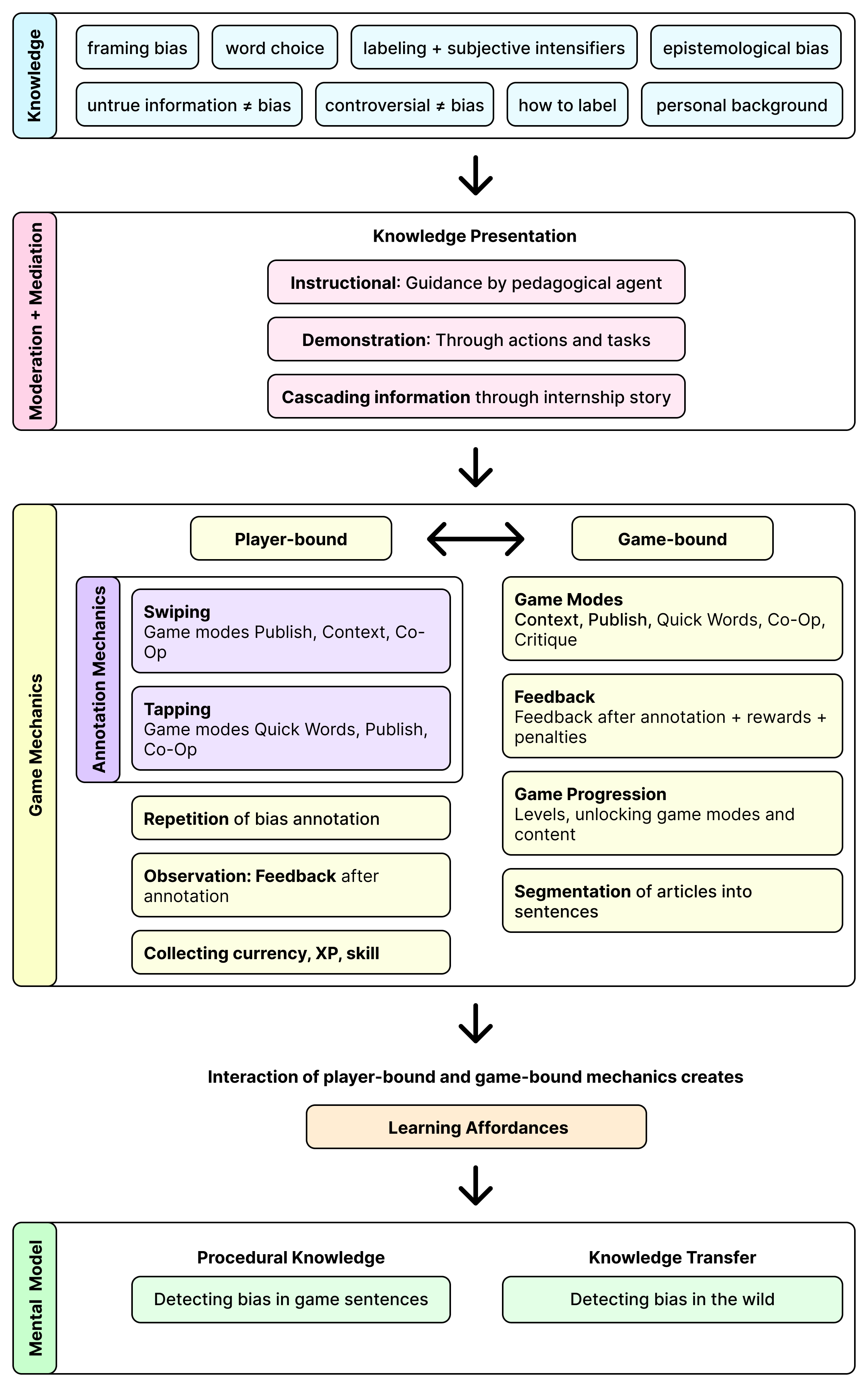}
 \caption{The figure shows how News Ninja applies the Gamified Knowledge Encoding Model to gamify bias learning and facilitate annotation through four steps. First, News Ninja breaks down learning objectives into knowledge units. Those units are presented to players through game mechanics like the pedagogical agent, demonstrations, or the narrative. Next, the interaction between player-bound and game-bound mechanics creates learning affordances. These allow players, through repetition, to form mental models and apply their knowledge.}
 \Description{A figure showing the interaction of game mechanics for learning, fun, and annotation. It is structured in four units: knowledge, moderation and mediation, game mechanics, and mental model. It starts with the units of knowledge for linguistic bias on top: framing bias, word choice, subjective intensifiers, epistemological bias, untrue information does not equal bias, controversial does not equal bias, how to annotate, and personal background. Then, moderation and mediation elements, which transfer this knowledge and its rules, are listed, with instructions by plant, demonstration through action and tasks, and cascading information in the story. Next, game mechanics are listed, with the interacting player-bound mechanics on the left and game-bound mechanics on the right. Player-bound mechanics include data mechanics swiping and tapping, with additional observation of feedback, repetition, and collection of skill, currency, and XP. Game-bound mechanics include the game modes, feedback, game progression, and content segmentation into sentences. Their interaction creates learning affordances, displayed below the game mechanics. From learning affordances, a mental bias model arises. This allows players to detect bias in new game sentences and, further, in the real world.}
 \label{fig:derivatives}
\end{figure}

The design process of News Ninja adapts the Gamified Knowledge Encoding Model \cite{oberdorferGamifiedKnowledgeEncoding2018} in a four-step process, detailed in Figure \ref{fig:derivatives}.
First, \textit{knowledge} is divided into short units, each covering a single learning objective.
Then, we fit game mechanics for the \textit{moderation and mediation} of the units.
The units' content is integrated through the pedagogical agent, demonstrating game modes, narrative, and repetition.
In the next step, we design player-bound and game-bound \textit{game mechanics} that allow for applying the new knowledge and frame the learning environment.
The annotation mechanic is integrated as a player-bound mechanic.
The interaction between game-bound and player-bound mechanics creates \textit{learning affordances} and enables the formation of \textit{mental models} through repeated interaction.
Such \textit{mental models} allow players to increase their game performance and detect bias in the real world.

Players start their journey, and the narrative introduces them as interns at a news outlet with an office plant as their pedagogical agent.
The plant explains why media bias is an important issue players can help with and why their personal background matters.
It then guides players through the demographic survey.
Next, the plant leads them through the interactive tutorial (Section \ref{sec:tutorial}) with immediate feedback (Section \ref{sec:feedback}) on the annotation mechanic (Section \ref{sec:mechanic}).
Before the main gameplay, players encounter previously classified sentences, reinforcing learning through direct feedback and assessing their bias detection skill.
Later, they unlock additional game modes and topics that incorporate social interactions and enable discussion.

\subsection{Data Annotation Mechanic} \label{sec:mechanic}
\begin{figure}[!htbp] \centering
 \includegraphics[width=\textwidth]{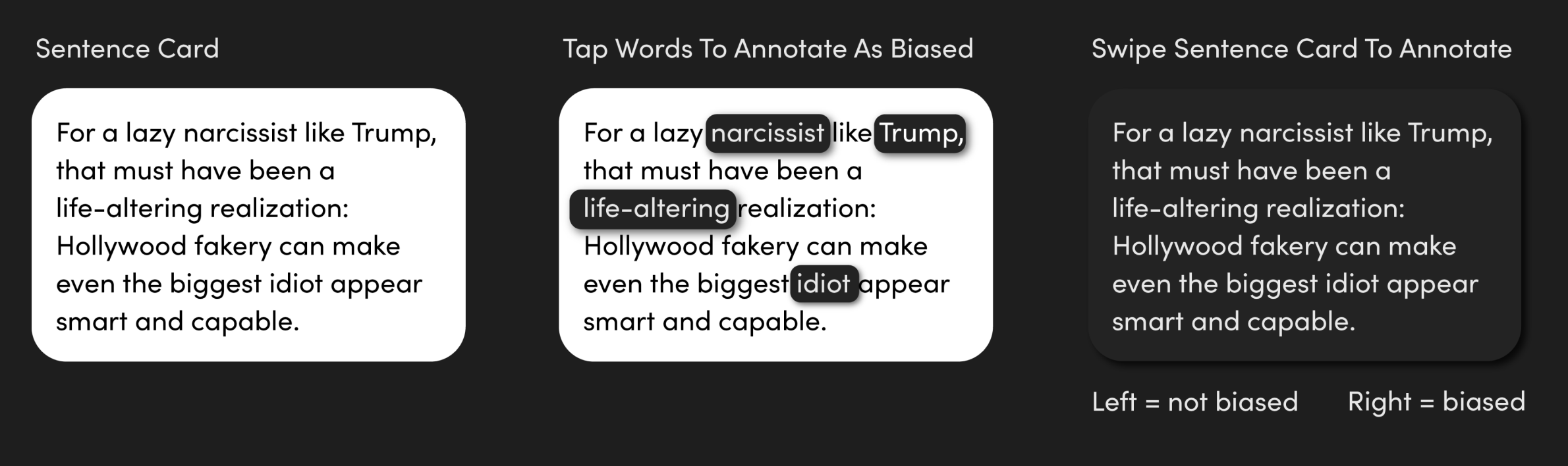}
 \caption{Interaction and data annotation mechanic of the \textit{Publish} game mode (Section \ref{sec:gamemodes:details}). First, players tap biased words. Then, they swipe or use the buttons to annotate the sentence as "biased" or "not biased."}
 \Description{Three sentence cards are depicted on a black background, with the first showing the unlabeled sentence, the second the mechanic of tapping biased words, and the third the mechanic of swiping the sentence card to annotate the whole sentence. The sentence reads, "For a lazy narcissist like Trump, that must have been a life-altering realization: Hollywood fakery can make even the biggest idiot appear smart and capable." On the second card, the words "narcissist," "Trump," "life-altering," and "idiot" are highlighted in black. This means a player annotated them as biased. The text below the third card states that a left swipe annotates a sentence as "not biased" and a right swipe as "biased."}
 \label{fig:GameMechanic}
\end{figure}

News Ninja divides data annotations into sentence level and word level mechanics.
This structure mirrors the structure of the BABE dataset \cite{spindeNeuralMediaBias2021}, a commonly used \cite{semeval} expert-level media bias dataset, which aims to cover linguistic bias at the lowest identifiable level and without the influence of article-level context.
In BABE, sentences are labeled "biased" or "not biased."
Each sentence can have biased words.
Hence, News Ninja's first application focuses on linguistic bias on the sentence level and excludes the article level, collecting binary bias annotations on word and sentence levels.
For sentence level annotation, pictured in Figure \ref{fig:GameMechanic}, a left swipe annotates a sentence as "not biased," while a right swipe annotates it as "biased."
Alternatively, players can use designated buttons.
For word level annotation, players select biased words by tapping on them.
Mimicking BABE, players of News Ninja can mark words as "biased" and still annotate a sentence as "not biased."

\subsection{Feedback} \label{sec:feedback}
\begin{figure}[!htbp] \centering
 \includegraphics[width=\textwidth]{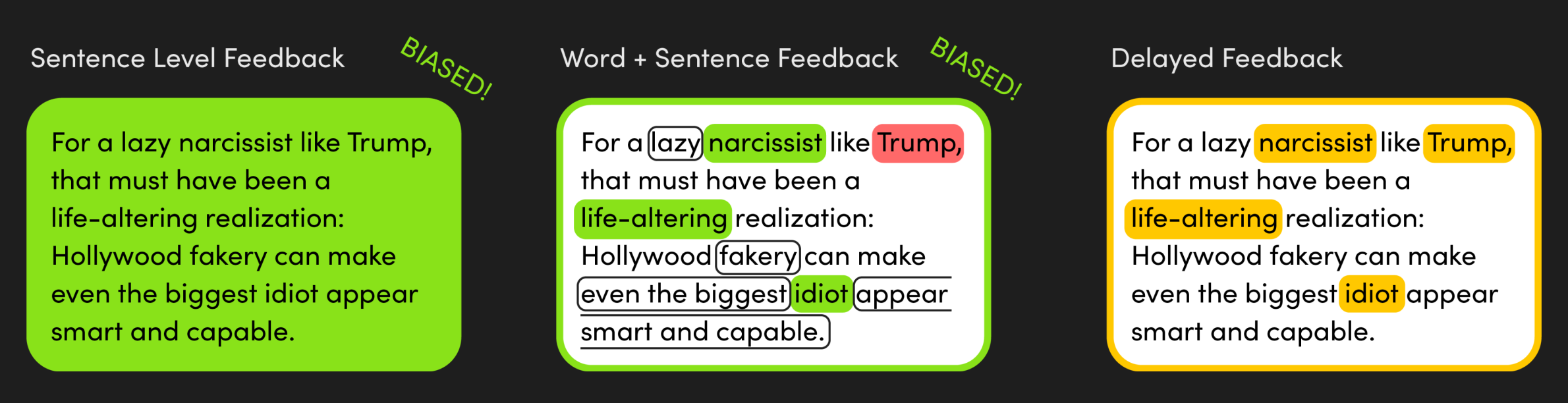}
 \caption{Direct and delayed feedback in the \textit{Publish} game mode (Section \ref{sec:gamemodes:details}). Sentence level feedback color-codes the card. In this case, it turns green and indicates a hit. In case of a miss, it turns red. Next, correctly and incorrectly annotated words are shown in green and red. Missed words display a black outline. Stopwords do not count in the feedback and are displayed as right when surrounded by biased words. The third card shows delayed feedback. If a ground truth has not been established, sentence and word level feedback is displayed in yellow.}
 \Description{Three sentence cards are depicted on a black background, with the first showing the feedback on sentence level, the second showing feedback on word level, and the third showing delayed feedback in case a ground truth has not formed. The sentence reads, "For a lazy narcissist like Trump, that must have been a life-altering realization: Hollywood fakery can make even the biggest idiot appear smart and capable." The first card has a green fill, indicating a hit on the ground truth. The second card displays a green outline, and the words "narcissist," "life-altering," and "idiot" are highlighted in green, also indicating hits. The word "Trump" is highlighted red, indicating a wrong bias annotation. The words "lazy," "fakery," "even the biggest," and "appear smart and capable" display a black outline, indicating them as missed biased words. The last card shows delayed feedback through a yellow sentence card outline and yellow highlights of the words "narcissist," "Trump," "life-altering," and "idiot."}
 \label{fig:Feedback}
\end{figure}

The game employs two types of feedback: direct feedback and delayed feedback (Figure \ref{fig:Feedback}).
Direct feedback is activated when the ground truth of a sentence or word is known.
Within this framework, "ground truth" refers to the label of a sentence or word.
A sentence level label is attained either when the sentence originates from BABE \cite{spindeNeuralMediaBias2021}, the baseline dataset,\footnote{Although BABE has a lower agreement score relative to other datasets, it is notably high for a media bias dataset, a reflection of the inherent subjectivity in media bias that complicates achieving consensus, especially at the sentence and word levels. Moreover, it is the most comprehensive dataset currently available to us. We discuss this in Section \ref{sec:dis:dataquality}.} or has at least five annotations by players who either voted "biased" or "not biased," thereby determining labels via majority vote \cite{Law2011}.\footnote{We further discuss mechanisms to determine labels in Section \ref{sec:dis:majorityvote}.}
Successful matches to the ground truth are rewarded with a green card outline, currency, skill, experience points, and sound effects; otherwise, the card turns red.
The ground truth label, biased or not biased, appears above the card (Figure \ref{fig:teaser}).
For word level bias, tapped words are highlighted in green or red, corresponding to the ground truth match (Figure \ref{fig:Feedback}).
The game considers a word "biased" when either two players (< 8 annotations) or 25\% of players marked it as biased.
Comparison against the ground truth and the hit/miss percentage calculation facilitates player rating, enabling the selection of inputs from players with higher bias detection rates.

However, for added sentences with no established ground truth, the "cold start problem" arises \cite{kicikogluWormingoTrueGamification2019} as the game cannot give direct feedback.
To navigate this challenge, News Ninja employs delayed feedback, visualized on the right in Figure \ref{fig:Feedback}.
Here, players receive feedback consisting of yellow visual cues indicating they can revisit the game at a future point when sufficient data is available.
The card outline turns yellow for delayed sentence level feedback, and a yellow dash icon appears.
On the word level, selected word cards turn yellow.
Then, the sentence moves to the \textit{Paper} section.
Players receive push notifications and see a yellow dot as a signifier on the navigation bar on the \textit{Paper} section icon when ground truth is established, indicating new information and rewards.
If players hit the ground truth, they receive a higher reward, while the uncertainty promotes extrinsic motivation \cite{chouActionableGamificationPoints2015}.
Players increasingly encounter unclassified sentences with delayed feedback as they progress and increase their detection skills.

\subsection{Tutorial} \label{sec:tutorial}
The tutorial progressively teaches linguistic bias through interactive examples and direct feedback while gradually increasing complexity at each level.
Each tutorial level encapsulates one to two learning objectives without using scientific terms.
Instead, the game aims for players to subconsciously learn to discern how bias manifests and identify it within sentence context.
Players see ten manually selected sentences to classify in each tutorial level while receiving immediate feedback to foster learning.
The tutorial starts with simple sentences and later transitions to more complicated ones.
Similarly, the completion of each level unlocks a new game mode.
The game modes incrementally increase the challenges by starting with the sentence level, progressing to the word level, and, ultimately, combining both.
As a game element that is part of the UI, the plant visually grows and blossoms with the player's progress, symbolizing their learning.

We base the tutorial's content on the annotation guidelines for the expert-annotated dataset by \textcite{spindeNeuralMediaBias2021}.\footnote{Derived on 20.07.23 from \url{https://github.com/Media-Bias-Group/Neural-Media-Bias-Detection-Using-Distant-Supervision-With-BABE/blob/main/annotation\_guidelines\_BABE.pdf}}
The tutorial mirrors these guidelines by covering the five elements of what media bias is, the importance of personal background, what needs to be annotated as biased, how to annotate, and what should not be annotated as biased.

The guidelines start with a general introduction to media bias, explaining that it can manifest through "particular word choice or framing exposing readers to non-neutral news reporting."
For instance, the term "Coronavirus" is presented as unbiased, while "Chinese Virus" is considered biased.
Further, it explains why it is important to be aware of linguistic bias and how it can manifest.
The game's objective is simultaneously made clear: to train players to read with greater critical awareness and contribute to detecting bias.

Then, the guidelines explain that personal background impacts bias perception; hence, a demographic survey is necessary.
News Ninja highlights the value of understanding players' viewpoints, followed by the demographic survey.
It further emphasizes that players should set aside personal opinions on any topic, regardless of its political implications or alignment with their beliefs.

Next, the guidelines outline various types of bias:
\begin{itemize}
  \item \textbf{Framing bias:} Skewing reader perception by only describing one point of view or frame.
  \item \textbf{Word choice:} Using one-sided terms or ideologically-driven depictions of concepts that alter readers' point of view.
  \item \textbf{Subjective intensiers}: Employing adjectives or adverbs that convey a strong opinion in that context, introducing bias.
  \item \textbf{Epistemological bias}: Manipulating language to affect the credibility of a statement, either enhancing or diminishing its believability.
\end{itemize} 
News Ninja adapts this structure as the plant illustrates how framing can sway readers' opinions by presenting events from a single viewpoint.
Subsequently, the plant discusses the impact of vague, dramatic, or sensational language and underscores how ambiguous or specific words can provoke emotional reactions.
While the guidelines detail the annotation process, News Ninja opts for a more hands-on approach and demonstrates the game mechanics directly to players.
Players are presented with a sentence, tasked with identifying biased words, and receive direct feedback (Section \ref{sec:feedback}).
Next, the tutorial continues with epistemological bias.

Upon level advancement, a new game mechanic is introduced, focusing on the sentence's entire context and how a topic can be controversial without containing bias.
The guidelines describe what should not be annotated as biased, stressing that controversial topics, or opinion-based reporting might not inherently be linguistically biased.
For example, "abortion" is not a biased word, but the term "abortionist" is biased.
The tutorial highlights that even statements containing untrue information do not necessarily feature biased language.

The final tutorial level explains the most complex annotation mechanic: Players first identify biased words by tapping and subsequently assess the entire sentence's bias using the swiping gesture or buttons.
This level lets players practice the annotation mechanism while learning from direct feedback.
Upon completion, it leads them back to the home screen.

\subsection{Game Modes and UI} \label{sec:gamemodes}
The game's home screen displays player statistics at the top, including in-game currency, experience points, and player level (Figure \ref{fig:teaser}).
A navigation bar on the bottom enables players to toggle between the home screen, the \textit{Paper} section, a repository of sentences, both previously played and awaiting feedback, the community section, and the shop.
The shop allows players to unlock new topics.
Central to the home screen is the \textit{Skill} bar and the \textit{Mission} bar, explained in Section \ref{sec:motivational}.
Moreover, the \textit{Breaking News} tile refreshes daily and showcases sentences in the \textit{Publish} game mode (s. Section \ref{sec:gamemodes:details}).
Below are the tiles for the five game modes.

\begin{figure}[!htbp] \centering
 \includegraphics[width=\textwidth]{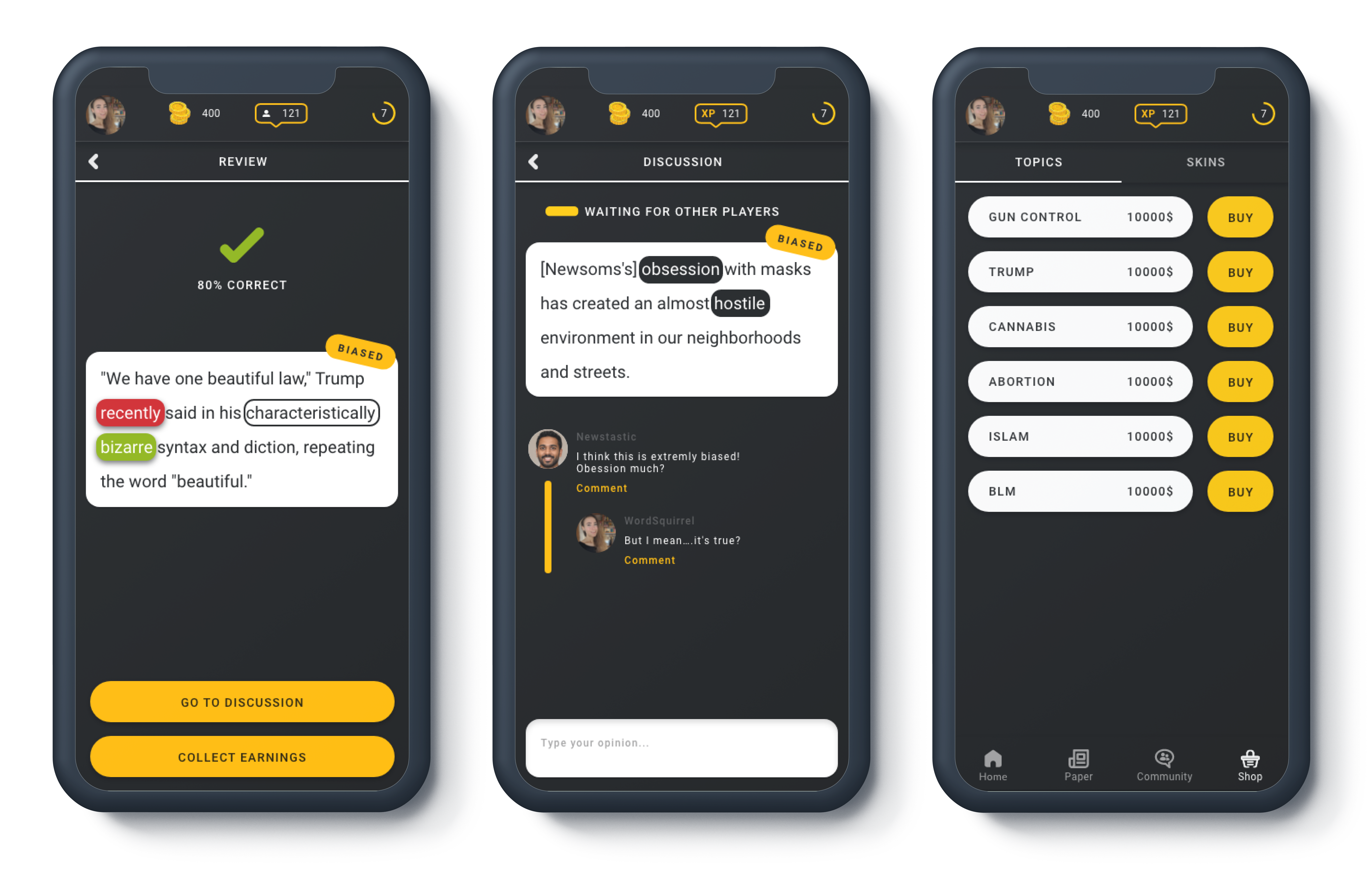}
 \caption{The first screen (left to right) shows the detailed feedback for one correctly annotated sentence in the \textit{Paper} section. It displays the sentence card with one word highlighted in red (incorrect annotation), one in green (correct annotation), and one with a black outline (missed word). Players can collect the reward or navigate to the discussion. The second screen shows the discussion of a sentence between two players. The sentence card shows on top with the comments below. The third screen shows the shop with six unlockable topics.}
 \Description{The first screen (left to right) shows the detailed feedback for one correctly annotated sentence in the \textit{Paper} section. It displays the sentence card with the sentence "We have on beautiful law, Trump recently said in his characteristically bizarre syntax and diction, repeating the word beautiful." "Recently" is highlighted in red (incorrect annotation), "bizarre" in green (correct annotation), and "characteristically" with a black outline (missed word). The text above states that the annotation that includes both levels was 80\% correct. Players collect the reward or go to the discussion with the two yellow buttons beneath. The second screen shows the discussion of the same sentence between two players. The sentence card shows on top with the comments below. The third screen shows the shop with six unlockable topics, in this case, "gun control," "Trump," "Cannabis," "abortion," "Islam," and "BLM."}
 \label{fig:UI}
\end{figure}

\subsubsection{Game Modes} \label{sec:gamemodes:details}
Effective bias detection relies on regular interaction with diverse content and feedback rather than pure theoretical understanding.
The five game modes (1) \textit{Context}, (2) \textit{Publish}, (3) \textit{Quick Words}, (4) \textit{Co-Op}, and (5) \textit{Critique} integrate the annotation mechanics differently to provide variety and cater to diverse player preferences.
They also foster a sense of progression and achievement by unlocking new game modes.
To increase fun, News Ninja introduces new elements into the mechanics, such as time constraints or cooperative challenges.

Post-tutorial, players can only access the (1) \textit{Context} and (2) \textit{Publish} game modes (Figure \ref{fig:teaser}).
Before playing, players select an available topic from which ten sentences are drawn.

The (1) \textit{Context} mode operates with the sentence level mechanic.
It shows a single sentence card to swipe with a \$10 virtual currency reward for matching the ground truth.
A left swipe annotates a sentence as "not biased," and a right swipe annotates it as "biased."
After ten sentences, the game provides a summary, showing correct and incorrect classifications and permitting sentence review.
News Ninja awards a bonus if players classify seven or more sentences correctly.

(2) \textit{Publish} operates on sentence and word level by combining both annotation mechanics (Figure \ref{fig:teaser}).
Players first identify biased words by tapping on them before swiping the sentence card as they do in \textit{Context}.
Feedback includes missed biased words highlighted with a black border (\ref{fig:Feedback}).
We count correct words as a bonus and do not punish misses as it is often hard to find all biased words.
Stopwords are automatically excluded from the calculation and shown as right if a biased word appears next to it.

The game assesses players before unlocking further game modes after the tutorial by presenting sentences with established ground truths and computing players' skill levels based on accuracy.

\begin{figure}[!htbp] \centering
 \includegraphics[width=\textwidth]{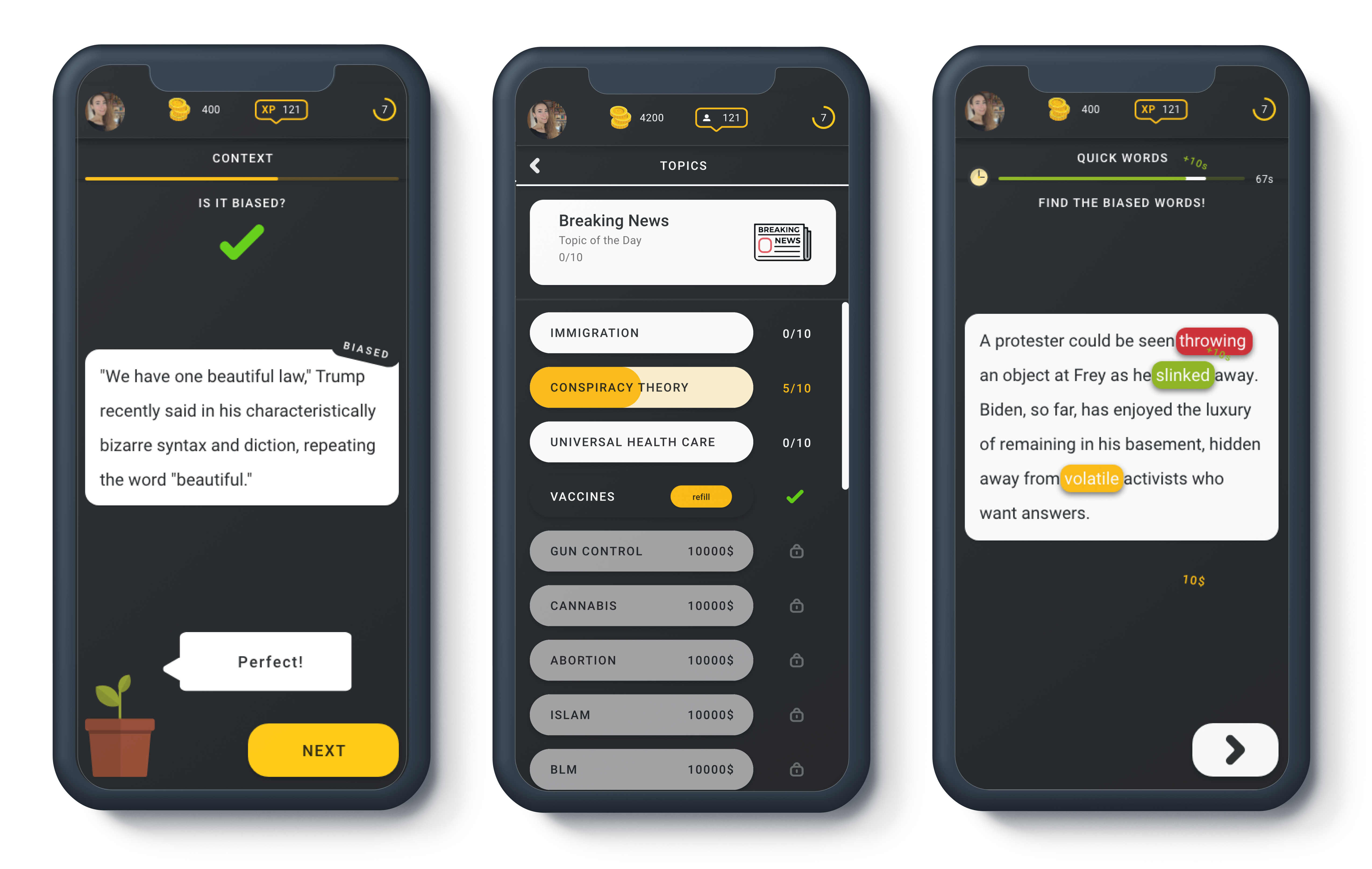}
 \caption{The first left screen shows feedback in the game mode \textit{Context}, with a sentence card displayed. It includes a green hook to indicate a correct answer on top, and the plant speaks motivationally in the left corner. The middle screen displays the topic selection before \textit{Publish}. One topic was already played (black button with yellow refill button), and three are playable (white buttons). The right screen displays a \textit{Quick Words} screen with the sentence card, feedback (green highlight for correct answer, red highlight for incorrect answer, yellow highlight for delayed feedback), the time bar, and the button for the next sentence in the bottom right corner.}
 \Description{The first left screen shows feedback in the game mode Context, with a sentence card displayed. The sentence card reads, "We have on beautiful law, Trump recently said in his characteristically bizarre syntax and diction, repeating the word beautiful." A green checkmark above indicates a correct answer, and the black biased label in the upper right corner of the sentence card shows that the player marked it as biased. The plant says "Perfect!" through a speech bubble in the left corner. The yellow "Next" button on its right lets players navigate to the next sentence. The middle screen displays the topic selection before Publish. The topic "Vaccine" was already played (black button with yellow refill button), and the three topics "Immigration," "Environment," and "Universal Health Care" are playable (white buttons). The topics "gun control," "Cannabis," "abortion," "Islam," and "BLM" are displayed in grey with a lock icon to their right as they are still locked. The right screen displays a Quick Words screen with the sentence card reading, "A protester could be seen throwing an object at Frey as he slinked away. Biden, so far, has enjoyed the luxury of remaining in his basement, hidden away from volatile activists who want answers." The word "throwing" is highlighted red (incorrect answer), "slinked" is highlighted green (correct answer), and "volatile" is highlighted yellow (delayed/ waiting for feedback). The time bar is on top, in green, with a white extension, showing that the last hit on "slinked" added 10 seconds. The button for the next sentence with a right-pointed arrow is in the bottom right corner.}
 \label{fig:GameModes}
\end{figure}

Next, the (3) \textit{Quick Words} mode is unlocked.
\textit{Quick Words} focuses on word level annotation and adds a timed challenge (Figure \ref{fig:GameModes}).
Players skip through sentences to tap as many biased words as possible before time runs out.
Correct classifications earn game currency and additional time.
Incorrect ones deduct time.
If there is no majority vote yet, yellow feedback tiles show.
A summary of identified words and their respective bias ratings shows when time runs out.
\textit{Quick Words} responds to research indicating greater word level than sentence level bias divergence among individual raters \cite{spindeNeuralMediaBias2021} and aims to increase word level annotations.
While there is a higher risk of biased judgments when making quick, automatic decisions \cite{sys1sys2InterventionsMoravec2020, AnchoringEffect1974}, News Ninja prioritizes fun and player engagement to later monitor data quality more closely (s. Section \ref{sec:dis:fw}).

The (4) \textit{Co-Op} mode allows cooperative gameplay and integrates both annotation mechanics.
Rewards are based on mutual agreement at both word and sentence classifications.
The faster player receives a bonus.
After the prior modes, we expect players to have achieved similar competencies, facilitating agreements.

The final (5) \textit{Critique} mode includes both annotation mechanics by showing prior player annotations.
Players can agree or disagree, adapt the ratings, and receive direct feedback when a ground truth forms (Figure \ref{fig:teaser}).
This game mode unlocks last because the game needs to ensure that players have collected enough experience to rate peers effectively.

\subsubsection{Paper Section}
Sentences from previous game rounds move to the \textit{Paper} section (Figure \ref{fig:teaser}).
This section provides players with an opportunity to reflect on their prior gameplay.
When a sentence with prior delayed feedback forms a ground truth, the game notifies players of the available feedback.
Collecting this feedback — when in alignment with the ground truth — yields greater rewards than direct feedback, incentivizing players to revisit the game with an additional element of surprise.

\subsection{Turning Player Input Into a Dataset}
\begin{figure}[!htbp] \centering
 \includegraphics[width=0.9\textwidth]{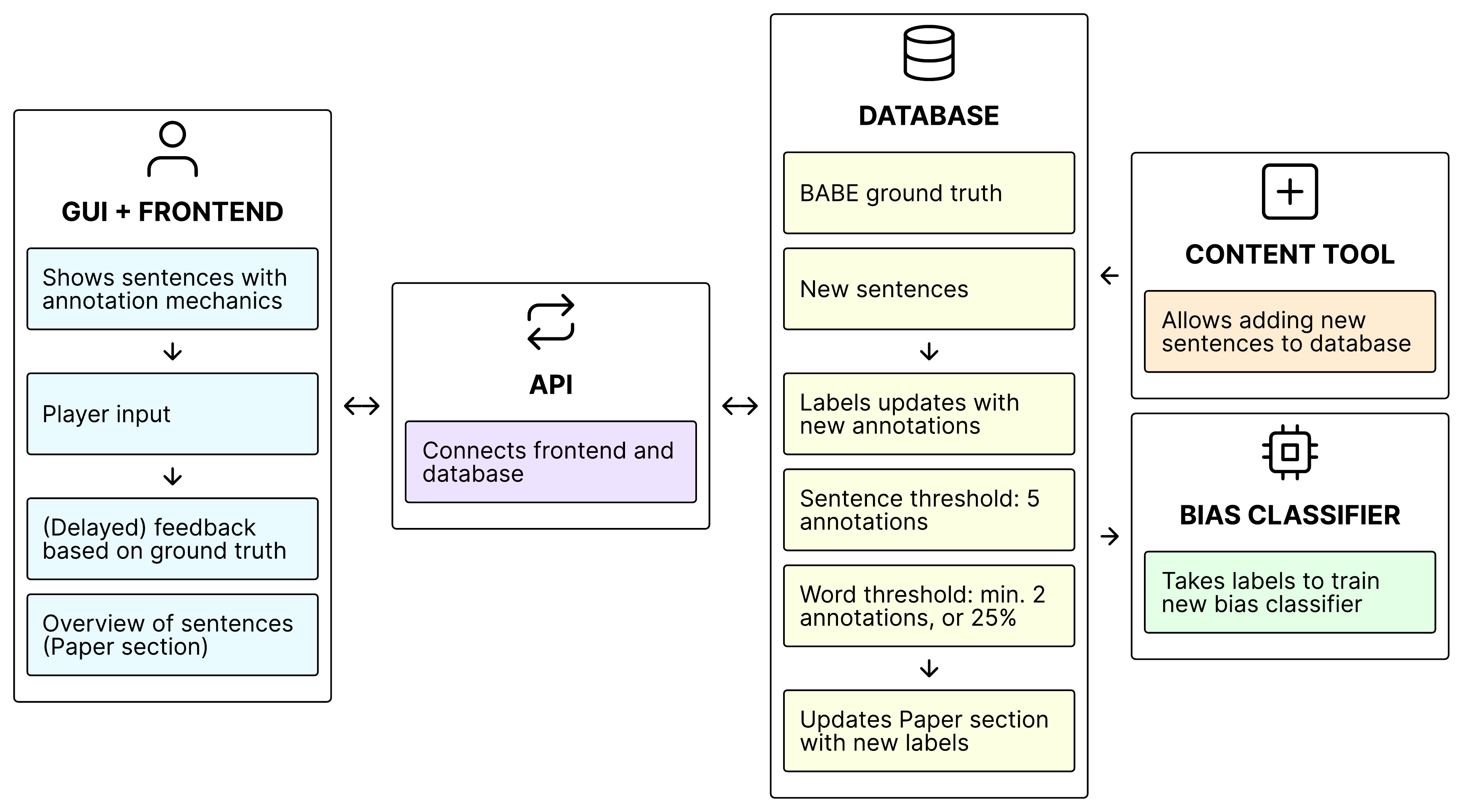}
 \caption{System architecture of News Ninja. The game collects annotations from players in the front end, updates the labels in the backend, connects through the API, and shows feedback based on the new labels. The last step trains a new classifier with the obtained labels.}
 \Description{The figure shows five blocks, starting on the left. First, the GUI and frontend block, then the API block, the Database block, and lastly, the Content Tool block above the Bias Classifier block. The GUI and frontend block contains four elements. The first block says, "Shows sentences with annotation mechanics. An arrow points down to the next block, "Player input." Another arrow connects it to "(Delayed) feedback based on ground truth" and "Overview of sentences (Paper Section)." The GUI and frontend block connects with the next right API block through an arrow pointing both ways. The single element in the API block says, "Connects frontend and database". To the right, the API block connects with the Database block through another double-pointed arrow. The Database block contains six elements. The first two elements, "BABE ground truth" and "New Sentences," connect with a downwards pointed arrow to "Labels updates with news annotations" and their thresholds, namely five annotations for sentences and two annotations or 25\% of annotators for words. Another arrow connects them to "Updates Paper section with new labels". Next, on the upper right, the Content Tool block points towards the database block with the element "Allows adding new sentences to database" inside. Below, the Database block connects to the Bias Classifier block with the element "Takes labels to train new bias classifier."}
 \label{fig:archi}
\end{figure}

To turn player annotations into labels, News Ninja accumulates them from the game modes described in Section \ref{sec:mechanic} on word and sentence level in the backend, as shown in Figure \ref{fig:archi}.
Once a sentence or word reaches its thresholds, we assign a bias label based on a majority vote \cite{Law2011}.
The resulting dataset contains sentence texts, sentence level labels, biased words within sentences, sentence topics, links to articles, the publishing outlet, and its respective leaning.
The system enables training new bias classifiers with the game dataset using the approach of \textcite{spindeNeuralMediaBias2021}.
In case of below-threshold annotations or a draw, players receive delayed feedback.
A word receives a bias label if identified as such by a minimum of two players or by 25\% of the players who encountered the sentence (Figure \ref{fig:archi}).
Due to the challenges of identifying bias at the word level and the lower agreement reported in prior research \cite{spindeHowWeRaise2022}, this threshold is deliberately set low.
Annotations are only collected once a player surpasses the tutorial levels to ensure a basic understanding of linguistic bias.
New sentences, including source and leaning in line with \textcite{spindeNeuralMediaBias2021}, are added via a web application designed for content integration for continuous updates.

\subsection{Motivational Game Elements} \label{sec:motivational}
News Ninja employs various motivational game elements to sustain player engagement, detailed in Table \ref{tab:allmechanics}.
A sense of \textit{progression} is achieved by buying new topics for sentences in the shop with in-game currency (Figure \ref{fig:UI}).
By collecting experience points, players progress through levels and unlock new game modes \cite{segundodiazBuildingBlocksCreating2022}.
Simultaneously, the plant grows with each level, symbolizing the players' growth.
A progress bar on the home screen visualizes players' skill levels --- players' accuracy on classified sentences --- to increase intrinsic motivation by monitoring one's learning process.
Beneath it, the group mission encourages a cooperative effort to mitigate bias by setting a number of labels as the goal that all players work on together, visualized through a progression bar below the skill level.

Daily, topics are refilled with limited sentences — capped at ten — to foster a feeling of \textit{scarcity}, which makes topics more desirable \cite{hamariGamificationMotivationsEffects2015}.
The cappping also ensures players annotate all ten sentences and choose to annotate from different topics \cite{chouActionableGamificationPoints2015}.
Players can purchase additional sentences using in-game currency (Figure \ref{fig:GameModes}).
Similarly, the \textit{Breaking News} tile on the home screen refills daily, offering higher rewards on daily completion (\textit{Publish} game mode) and encouraging a build-up of consecutive play streaks.
Further, it allows the developers to inject and prioritize sentences to balance the dataset selectively.
Delayed feedback incentivizes players to return and collect rewards, reinforcing recurring player interaction through \textit{unpredictability}.
\textit{Social motivation} is introduced through sentence discussions with other players (Figure \ref{fig:UI}).

\subsection{Ethical considerations} \label{sec:ethics}
The primary objective of the News Ninja game is to educate players about linguistic bias to collect bias labels through a gameful approach.
While data collection is a significant aspect of this endeavor, we consciously avoid employing manipulative game design patterns \cite{tuiteGWAPsGamesProblem2014, grayDarkPatternsSide2018b} aimed at coercing players to annotate more extensively.
Instead, the game leverages players' intrinsic motivation to learn and contribute to mitigating bias, even at the expense of compromising the amount of data collected.
Streak loss, the unpredictability of delayed feedback, and the capping of sentences are the only design patterns we consider more influencing in News Ninja.

\section{Study Design} \label{sec:studydesign} \label{sec:studydesignFINAL}
Initially, we conducted a qualitative pre-test of the tutorial (Q1) and the annotation mechanic (Q2) to assess players' experience (Section \ref{sec:pretest}) and refine the game design (Section \ref{sec:game}).
We moved on to the primary study with crowdsourced workers to evaluate whether player-generated data achieves similar quality as expert labels (Q3, Section \ref{sec:studydesign:outline}).

\subsection{Pre-Test of Tutorial and Data Annotation Mechanic} \label{sec:pretest}
To evaluate players' experience with the tutorial (Q1) and annotation mechanics (Q2), we conduct an A/B test with a UX survey with 21 participants on an early News Ninja mobile version.
The game group played the game's tutorial, and the control group read through an introduction to biased wording used in prior studies \cite{spindeTASSYTextAnnotation2021} before annotating the same 20 sentences in the \textit{Publish} game mode.
The study aims to identify potential design flaws, assess whether the tutorial and game mechanics are enjoyable, understandable, and easy to use, and determine player motivation, likes, and dislikes.
The UX survey incorporated the Single Ease Question (SEQ) \cite{sauroComparisonThreeOnequestion2009} to query task difficulty, the 20-item Intrinsic Motivation Inventory (IMI) \cite{mcauleyPsychometricPropertiesIntrinsic1989} to assess player motivation, and three open-ended questions regarding first impressions, general experience within the game, and any encountered problems (Section \ref{sec:app:uxquestions}).
To evaluate differences in performance, we compare player annotations against the gold standard data set by \textcite{spindeNeuralMediaBias2021}.
As our goal is to understand the experience of the game group more comprehensively, and since prior studies already used the control group's bias introduction, we randomly assigned 15 players to the game group and 6 to the control group.

We recruit volunteer participants through university group mail.
Sixteen participants were between 20 and 29 years old, and five were between 30 and 40.
One identified as women, 18 as men, and two as diverse.
Regarding educational background, 14 held a bachelor's degree, five had completed graduate work, one had finished high school, and one had partially completed high school.
Regarding language proficiency, 14 participants were fluent in English, and seven were at an intermediate level.
The political orientation of the sample showed a left slant; 12 participants identified as leaning left, three as leaning right, and six as centrist.
As for media consumption habits, 3 participants consumed news several times per day, six daily, nine several times per week, one several times per month, and two rarely.

The analysis revealed that the game group ($n = 15$) had a 8\% greater alignment with the gold standard ($M_{\text{Game}} = .81, SD = .11; M_{\text{Control}} = .75, SD = .06$) compared to the control group ($n = 6$), suggesting the tutorial enhanced annotation quality.
However, the increase is insignificant due to the small sample size.
Feedback from the game group underscored a broad appreciation for the game's approach to media bias.
Participants especially valued the guidance provided by the plant.
The game group had a mean SEQ of 4.6/5 ($SD = .34$), and the control group had a mean of 3.6/5 ($SD = .87$).
The IMI scored 5.2/7 ($SD = .54$) for the game group and 4.6/7 ($SD = .92$) for the control group.
In line with the SEQ results, control group participants reported feeling overwhelmed by the task, indicating that the annotation guidelines from previous studies may have been too ambiguous.
The open-question answers also highlighted a high complexity in the annotation mechanic.
Other feedback expressed a wish to revisit prior annotations and pointed to the potential advantages of additional gamification and narrative integration.
In response to the complexity noted by players, we refined the tutorial by simplifying the wording, shortening each lesson, and dividing the lessons into smaller sections followed by interactive examples.
To simplify the game mechanics, we segmented them into five distinct game modes, described in Section \ref{sec:gamemodes}, and added a section to review past and delayed sentences.
Next, we develop the redesign as a mobile-first web application that does not require the installation of an app.

\subsection{Study Outline} \label{sec:studydesign:outline}
The final study takes participants through a streamlined game version, aiming to ensure comparability by minimizing potential other variables that could influence player behavior.
The goal is to re-annotate and analyze the data quality of 10\% of BABE sentences (370 sentences) and 150 new sentences, resulting in 520 sentences that each need at least five player annotations (2600+ annotations).
To keep the study duration around 20 minutes, each player must play through the tutorial and 30 randomly and equally distributed sentences.
Hence, the study requires 100 participants ($100$ Players $* 3$ rounds $* 10$ Sentences $= 3000$ Annotations) that we recruit from the US on the micro-tasking platform Prolific with a payment of 6£ per hour.

Participants began by reading the data processing agreement.
On agreement, they continued to the game; otherwise, they were redirected to Prolific.
Participants progress through the demographic survey\footnote{As we conduct the study on Prolific, participants agree to complete the entire survey. Future online versions will offer the option to skip each question.} through which we assess players' backgrounds to monitor dataset bias.
The survey includes questions on age, gender, nationality, education, political leaning, news consumption frequency, and English proficiency (Section \ref{sec:app:demo}).
Political orientation is important for assessing bias through slant, and English proficiency is essential for grasping linguistic nuances like bias.

The game has three phases.
In phase 1 (tutorial), players learn about different types of linguistic bias and the game mechanics through the interactive tutorial described in Section \ref{sec:tutorial}.
In phase 2 (direct feedback), players play 20 randomly drawn sentences in the \textit{Publish} game mode (Section \ref{sec:gamemodes}) and receive direct feedback.
In phase 3 (delayed feedback), players saw ten new sentences with delayed feedback.
Upon completion, participants are thanked for their contribution and guided back to Prolific for payment.

We use Krippendorff's $\alpha$ as the Inter-Annotator Agreement (IAA) metric in our initial data quality assessment.
IAA measures the consensus among annotators on labeling tasks beyond what would be expected by chance alone.
This metric is widely recognized for its reliability in evaluating dataset reliability \cite{hayesAnsweringCallStandard2007} and is frequently used to analyze linguistic and media bias datasets \cite{spindeIntroducingMediaBias2022}.
We benchmark the IAA of News Ninja against the IAAs of two datasets within the domain of media bias.
Firstly, we compare it to the crowdsourced dataset MBIC \cite{spindeMBICMediaBias2021}, generated by non-experts who received a textual introduction to media bias.
This comparison is particularly relevant due to the similar recruitment methods through microtasking platforms.
Secondly, we compare News Ninja's IAA with the currently most extensive, expert-curated dataset BABE, developed by students and researchers focusing on media bias \cite{spindeNeuralMediaBias2021}.

Since IAA only measures agreement, we analyze the 370 re-labeled BABE sentences and the 150 new sentences by comparing them to newly created expert labels.
Our objective is to assess the degree to which player labels align with expert labels, in extension determining the effectiveness of the tutorial.
We further manually assess the types of sentences where players diverge from experts.
In addition, we compare the original BABE labels with the new expert labels \cite{spindeNeuralMediaBias2021} to evaluate the suitability of BABE as a ground truth for player training.

\subsection{Material}
The BABE dataset \cite{spindeNeuralMediaBias2021} functions as both the evaluation benchmark and the ground truth for player training.
To ensure comparability, participants re-annotate 10\% of the original dataset (370 sentences).
As relying solely on one dataset might add bias and the game's objective extends beyond mere re-labeling to create an extensive, crowdsourced linguistic bias dataset, two researchers compiled 150 new unlabeled sentences.
Hence, we test if the system can generate new labels with sufficient quality.
We limit the number of new sentences to 150 to keep the game duration around 20 minutes.
All sentences, including those from the open-source dataset BABE, are sourced from publicly available data on news websites.
The collection used the topics and timeframe of \textcite{spindeNeuralMediaBias2021} and leveraged AllSides\footnote{\url{https://www.allsides.com/}} to ensure balanced political representation.
Similar to \textcite{spindeNeuralMediaBias2021}, the ratio of "biased" to "not biased" sentences is 2:1.
Therefore, two datasets emerge from the study design: one with re-annotation sentences and one with new sentences.

\subsection{Participants and Inclusion Criteria}
The study involved 100 Prolific-recruited participants.
We briefed prospective participants on the study's details, estimated duration, and compensation via Prolific.
Those interested navigated to the React game application within their browsers (mobile or desktop).
The platform's design ensured comprehensive data collection upon game completion.
Thus, successful game completion served as the primary inclusion criterion.
Additionally, participants self-reporting English proficiency below an intermediate level were excluded.
We preliminary evaluated the study platform to test Prolific integration and data processing accuracy.

\section{Results} \label{sec:results}
\subsection{Demographics} \label{sec:demographics}
Participants had an average age of 36.72 years.
Of all participants, 50\% identified as men, 47\% as women, and 3\% as diverse.
Every participant was of US nationality.
34\% held a Bachelor's degree, 22\% had some college education, 21\% had completed high school, 12\% had pursued graduate studies, 8\% held an associate degree, 2\% had undergone vocational or technical training, and 1\% chose not to disclose their educational background.
Political orientation was assessed with a 21-point scale (0 = left, 20 = right).
Participants showed a left slant with an average score of 6.97.
Most participants were frequent news consumers.\footnote{The types of news media, including digital, print, or TV, were not specified.}
31\% consumed news daily, 29\% multiple times a day, and 23\% several times a week.
Only 13\% consumed news a few times a month, and 4\% either never consumed news or did so very infrequently.
Four participants reported intermediate English proficiency; all others indicated advanced proficiency.
Thus, we exclude no participants based on proficiency.
The average completion time was 21.45 minutes, with a median of 20 minutes.

\subsection{IAA Assessment} \label{sec:resultsdata}
\begin{figure}[!htbp] \centering
 \includegraphics[width=0.6\textwidth]{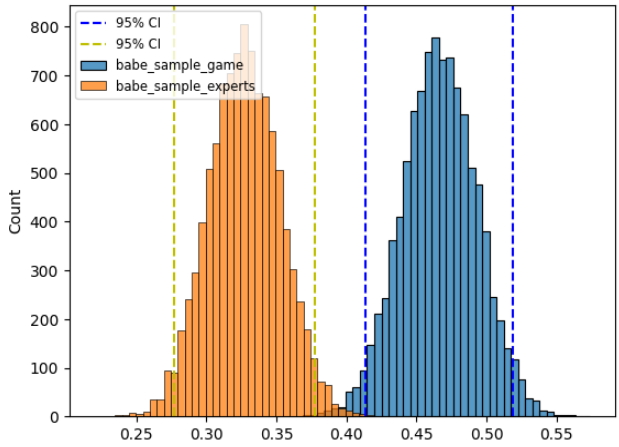}
 \caption{Bootstrapped histogram of Krippendorff's $\alpha$ for game-annotated (blue) and expert-annotated (orange) datasets. The two confidence intervals do not overlap, indicating a significant increase in IAA.}
 \Description{Histogram of Krippendorff's $\alpha$ for bootstrapped game-annotated and expert-annotated datasets. The two bootstrapped confidence intervals of the game-annotated dataset (blue) and the expert-annotated dataset (orange) do not overlap, indicating a significant increase in IAA.}
 \label{fig:sig}
\end{figure}

The sentences extracted from the BABE dataset and annotated with the game interface achieve an IAA (Krippendorff's $\alpha$) of 0.44.
This IAA surpasses similar crowdsourced annotations, which reported $\alpha$ = 0.21 \cite{spindeMBICMediaBias2021} — an increase of 109.52\%.
Moreover, it outperforms the expert annotations that recorded a Krippendorff's $\alpha$ of 0.39 \cite{spindeNeuralMediaBias2021}, an increase of 10.28\%.
Figure \ref{fig:sig} shows a histogram of the bootstrapped game-annotated dataset (blue) and the expert dataset (orange).
The 95\% confidence intervals of the two datasets do not overlap, indicating a significant increase in Krippendorff's $\alpha$ between the two datasets.

The new set of sentences achieves Krippendorff's $\alpha$ of 0.399.
This figure matches the IAA of expert annotations \cite{spindeNeuralMediaBias2021} and represents a 90\% increase compared to crowdsourced annotations \cite{spindeMBICMediaBias2021}.
The significant increase in IAA indicates that News Ninja could generate labels on new sentences that were comparable to expert quality in terms of annotator agreement.

\subsection{Comparison to Expert Labels} \label{sec:resultsdataexperts}
The player sentence labels achieve an accuracy of 79.8\% with the new expert labels, with a precision of 95.5\% and a recall of 69.2\%.
Players were more prone to missing actual positives but could reliably avoid false positives (Table \ref{tab:confusion}).
From the 370 re-labeled BABE sentences, 289 player labels match the expert ratings, while 81 differ, resulting in an agreement rate of 78.11\%.
In analyzing the sentences with discrepancies, we find that 74 sentences (91.36\%) identified as biased by experts were not labeled as biased by players.
A manual analysis reveals that these sentences typically feature subtle linguistic bias, characterized by one or two bias-inducing words.
For instance, the experts assigned a "biased" label to the following sentence, while players did not: "At the time, Wilson claimed he was only helping illuminate the contours of the 2016 race, before eventually being hired by one of the White House hopefuls."
Here, the word "claimed" introduces bias by questioning the believability of Wilson \cite{Recasens_2013}.

Conversely, only seven sentences (8.64\%) labeled unbiased by experts were labeled as biased by players.
These sentences frequently involve controversial topics such as transgender issues, racial issues, abortion and pregnancy, immigration, or vaccines.
In two cases, direct quotes contained strong words.
While the quote is biased, the sentence reporting it is not biased.
A comparison between player labels and BABE labels shows a 79.13\% match, with 77 sentences receiving differing labels.
The concordance between BABE and expert labels mirrors that of player and expert labels, standing at 61 sentences with an agreement of 83.47\%.
In some cases, player labels correct the labels from BABE.
For instance, the sentence "And, of course, along the way, the hipster CEO is absolved of any responsibility to help uphold good cultural order." is labeled as "not biased" in BABE.
However, both game labels and the new expert labels identify this sentence as biased.

\begin{table}
\caption{Confusion matrix of News Ninja labels and expert labels.}
\label{tab:confusion}
\begin{tabular}{>{\raggedright}p{2.7cm}cc} 
\toprule
 & \textbf{Player Positive} & \textbf{Player Negative} \\
 \midrule
\textbf{Expert Positive} & 213 & 95 \\
\textbf{Expert Negative} & 10 & 202\\
\bottomrule
\end{tabular}
\end{table}

Analyzing the 150 newly added sentences, we find an agreement of 84\% with the expert standard.
126 sentences were labeled correctly, while 24 differed.
Of the differing sentences, 21 were labeled as not biased (87.5\%), while three were labeled as biased (12.5\%).

\section{Discussion} \label{sec:dis}
This paper presents News Ninja's design and first evaluation, a GWAP designed for linguistic bias education and data collection, potentially heightening players' bias awareness during future news reading.
This first iteration tests data collection with annotation mechanics for crowdsourcing subjective truths, demonstrating that News Ninja achieves IAA levels and data quality comparable to expert datasets, suggesting its viability for creating linguistic bias datasets (Section \ref{sec:resultsdata} and Section \ref{sec:resultsdataexperts}).
While isolating the effects of single game mechanics is challenging, we surmise that the interactive tutorial and annotation mechanics significantly increased the IAA through the immediate feedback.
An extensive summary of News Ninja's game mechanics and their potential effects can be found in Table \ref{tab:allmechanics}.
The game-centric system is scalable and promising to be cost-effective by leveraging crowdsourced players over experts \cite{spindeAutomatedIdentificationBias2021}, 
It mitigates the risk of dataset obsolescence by periodically updating contents to capture changes in news, context, and perception over time.

\subsection{Guidelines to Tutorial (Q1)}
Distinct from prior research on media bias annotation, our study participants received instant feedback after written guidelines that fosters direct learning from their input.
The tutorial possibly contributed to the higher IAA and agreement with the expert standard through the clear structure, storyline, pedagogical agent, and broken-down learning objectives.
While we aim to keep the explanations as brief as possible, there is a risk that players may quickly skim or click through the text.
Consequently, it is essential to test players after the tutorial and establish a baseline for their bias detection skills, which can be incorporated into models generating the bias labels.
Even if players skip through the text, the feedback provided during the tutorial might contribute to their learning.
However, as this study primarily focuses on evaluating the quality of player labels, learning effects must be examined separately (Section \ref{sec:dis:fw}).

As the tutorial sentences are manually selected, they are straightforward examples that remain relevant over time.
However, the game must undergo constant review to adapt to changes in language.
We expect the topics and content to change faster than the language and expressions of bias.
Concurrently, we must ensure their relevance by periodically reviewing the teaching content.
Involving educational scientists in later game iterations will further support this aim.

\subsection{Annotations as a Game Mechanic (Q2)}
Preliminary testing revealed that some players found the initial \textit{Publish} game mechanics complicated.
Hence, the tutorial redesign introduces two game modes to prepare players.
The increased IAA of the quantitative study at both word and sentence levels suggests its potential success.
The exact impact of individual mechanics remains elusive, and isolated testing might be unproductive as they are integral to the game.
Despite acknowledging the topic's importance, some players perceived the annotation task as work-like (Section \ref{sec:pretest}).
This raises concerns about motivation without financial incentives, necessitating a separate study to evaluate the player experience.

\subsection{Data Quality (Q3)} \label{sec:dis:dataquality}
In terms of IAA, the re-annotated dataset outperformed the expert dataset \cite{spindeAutomatedIdentificationBias2021} by achieving a 10.28\% higher IAA than BABE and a 109.52\% higher IAA than the crowdsourced dataset.
This speaks for the advantage of combining annotation mechanics with game-based learning, gamification, and a feedback system \cite{kirschnerTenStepsComplex2017, oberdorferBetterLearningGaming2021} over traditional data annotation methods like Excel tables or annotation tools, possibly resulting in high consistency among participants.
We believe the game's approach makes understanding the task easier and more engaging than reading through annotation guidelines, which we will investigate in future work (Section \ref{sec:dis:fw}).

As observed in Section \ref{sec:resultsdataexperts}, players trained with BABE sentences reproduced the BABE annotations with 79.13\% agreement, indicating the training process's effectiveness.
However, this also raises concerns regarding the necessity for well-balanced training data to ensure proper performance and prevent the reproduction of biases within the training material.
We regard BABE as the ground truth, a notion that is inherently problematic given the subjective nature of bias and the questionable premise of a singular, definitive ground truth.
BABE may fall short of achieving this quality, as seen in the discrepancy of 16.53\% in expert ratings.
The dataset mostly mislabels sentences as "not biased" that contain subtle linguistic bias.
In addition to manually selecting tutorial sentences for future iterations, the sentences for all first six levels should undergo manual selection.
Similarly, increasing the threshold for sentence label decisions could ensure that players receive precise feedback.

We consistently saw that players recognized sentences with high levels of bias as biased.
Conversely, sentences with low levels of bias were often overlooked and not labeled as biased (Section \ref{sec:resultsdataexperts}).
This issue may stem from the binary labeling system currently in place.
When faced with uncertainty over whether content is biased, players might be inclined to categorize sentences as unbiased.
Introducing a scale could reflect these nuances, although it requires adjustments to the game's mechanics and feedback system, particularly regarding rewards.

Moreover, News Ninja should closely monitor the labeling of sentences on topics commonly perceived as biased, including transgender rights, race, religion, women's rights, and queer issues.
Previous NLP GWAP datasets contained stereotypes regarding gender and sexual orientation \cite{otterbacherCrowdsourcingStereotypesLinguistic2015}.
In News Ninja, experts could check sentences involving controversial topics or subtle biases, similar to the system described by \textcite{Demartini2020HumanintheloopAI}.
One solution to further balance the dataset would be to oversample sentences with low bias content with the help of experts.

\subsection{Cognitive Bias and Cultural Truth} \label{sec:dis:majorityvote}
News Ninja is the first GWAP to address the challenge of linguistic bias data creation.
In this section, we present three possible influences on data quality: (1) Bias through agreement, (2) cognitive biases, and the implications of (3) cultural truths.
The goal of building better classifiers for bias detection strongly influences News Ninja.
It aims to enhance agreement (IAA) for classifier training, resulting in a conflict between increasing IAA and diversifying opinions (1).
Hence, the design could increase convergence between player annotations, mainly through the tutorial and the feedback based on BABE's ground truth, possibly leading to dataset bias by reinforcing BABE annotation patterns.

Cognitive biases (2) can introduce potential irrationalities or deviations from normative decision-making processes \cite{AnchoringEffect1974} while annotating content for media bias.
Especially during crowdsourcing tasks where the objective determination of "true" answers is often elusive \cite{aroyoCrowdsourcingSubjectiveTasks2019}, they can decrease the quality of the annotated data \cite{hubeUnderstandingMitigatingWorker2019}.
To address this problem, \textcite{drawsChecklistCombatCognitive2021} propose a checklist to identify and mitigate cognitive biases in crowdsourcing tasks.
Within News Ninja, we identified four specific biases that could potentially compromise the quality of our dataset:
\begin{enumerate}
    \item Self-interest bias, which may incline annotators to skew data in favor of their political beliefs or inattentively overlook subtle biases while prioritizing speed over accuracy.
    \item Groupthink bias can be reinforced by displaying majority votes to players, thus encouraging conformity.
    \item Availability bias might affect judgments through the preconception of stereotypes within sentences.
    \item The anchoring effect \cite{AnchoringEffect1974}, potentially introduced via tutorial content that reflects BABE's ground truth, may also influence annotators' judgments.
\end{enumerate}

Further, agreement on linguistic bias is a cultural truth (3), the consensus based on a group of people's beliefs --- a judgment based on perception rather than an objective truth (e.g., if a word is a noun) \cite{Law2011}.
We assume that there is a cultural agreement between people of the same group that can be identified by aggregation, despite individual deviations \cite{romneyCultureConsensusTheory1986}.
However, linguistic bias, political leaning, and cultural backgrounds separate players into different groups.
Hence, a mere "ground truth" by a majority vote may be inadequate.
Achieving 100\% accuracy or agreement may be unrealistic, especially considering the evolving nature of language and meanings.
A dynamically evolving dataset might capture these changes more accurately.
Even experts might diverge, as seen in BABE's IAA, or bias the dataset, given their shared academic backgrounds.
Therefore, we suggest educating diverse annotators about media bias to capture a societal average.
Classifier results should be viewed as pointers, enabling analysis and augmenting the cognitive capabilities of news readers rather than replacing them.

We propose the following approaches to address the three challenges.
First, we implemented a check through manual expert analysis and a comparative assessment of the annotated data (Section \ref{sec:resultsdataexperts}).
Additionally, we integrated querying demographic metrics, such as political leanings, into the game \cite{drawsChecklistCombatCognitive2021}.
News Ninja must ensure the inclusion of data from players exhibiting unique annotation patterns or perspectives \cite{Law2011, spindeMBICMediaBias2021}.
Hence, a more sophisticated latent class model could account for the noisy data caused by individual biases such as political ideology, cultural backgrounds, possible gender effects, task difficulty, or expertise \cite{NIPS2010_0f9cafd0, yanActiveLearningCrowds2011}.
Incorporating these metrics into a probabilistic model and applying statistical hypothesis testing could identify systematic patterns that indicate the presence of cognitive biases \cite{drawsChecklistCombatCognitive2021}.

Certain design decisions require revisions to better align with our objective of developing a dataset for model training that includes diverse perspectives.
The current version is agreement-oriented.
As discussed, striving solely for consensus may not effectively capture cultural truths.
Consequently, we are transitioning to a process-oriented game design.
This approach rewards player actions and sustained participation rather than consensus per se.
\textit{Co-op} enhance agreement between raters by rewarding the replication of existing annotation patterns.
Instead of rewarding agreement, it could reward the annotation and provide players with a comparison afterward, potentially revealing the political leanings of their counterparts.
Presently, skill levels are determined based on agreement with the established ground truth.
Instead, News Ninja could incorporate specific test sentences with more objectively identifiable instances of bias for skill assessment purposes \cite{drawsChecklistCombatCognitive2021}.
Additionally, experience, mainly measured by the number of annotations, could count toward skill level.
Further, designing non-binary, more flexible annotation mechanics could increase diversification and accommodate the multifaceted nature of bias \cite{spindeIntroducingMediaBias2022}.
Subsequent analyses should determine the legitimacy of such contributions and develop approaches for their management.

\subsection{Limitations}
A significant limitation to consider is that monetary compensation was the predominant motivation for participation.
This leads to uncertainties about the game's genuine appeal without financial incentives and is further addressed in Section \ref{sec:dis:fw}.
This study focused on data quality by comparing player and expert labels.
However, we did not assess whether the tutorial and exercises improved players' bias detection skills.
While we compared their labels to expert labels, we did not test their bias detection skills before playing through the tutorial.
Consequently, players' long-term and short-term learning outcomes must be evaluated in a separate study, as detailed in Section \ref{sec:dis:fw}.

The research team's Western-centric bias and the dataset bias from \textcite{spindeAutomatedIdentificationBias2021} may manifest in the News Ninja datasets.
This potential influence could create a self-reinforcing loop, as players were given feedback from a dataset exhibiting similar biases \cite{spindeNeuralMediaBias2021}.
Further, segmenting articles into statements and sentences is limiting, as they show up without the context of the whole article.
The game, tested in a constrained and streamlined version, may only partially represent the full gameplay experience.
As we focused on the data quality after the tutorial in this iteration, we did not employ other gameplay metrics or survey instruments.

While the participant pool demonstrated gender diversity, it lacked a balanced representation in age, education, and nationality, potentially skewing representation and elevating IAA scores.
Moreover, the participants primarily came from Prolific, a platform recognized for its Western academic leanings.
Hence, the demographics of the players who generated the annotations must be considered to avoid bias in the dataset.

Likewise, the preliminary UX study (Section \ref{sec:pretest}), while offering insights into the complexity of the first game mechanic, has limitations.
The increase in agreement with the expert standard among the game group is not significant, and the groups were unevenly distributed, with significantly more players in the game group.
While focusing on the game group allowed for more qualitative insights, it limited the study's quantitative explanatory power.
Nearly all participants identified as men and had a high level of education.
Although the study identified and subsequently addressed some design flaws, the UX of News Ninja requires a more comprehensive analysis, as detailed in the following section.

\subsection{Future Work} \label{sec:dis:fw}
The following four next steps guide future research:

\begin{enumerate}
  \item The game offers potential as an educational tool, especially within academic settings. In our next step, we will study learning effects by conducting tests pre- and post-gameplay. We will further focus on extending the tutorial in cooperation with educational researchers and testing it with a student audience. This step includes manually re-selecting sentences for the first six levels. We focus on subtle linguistic bias in later lessons to resolve the issues with imprecise BABE labels and help players achieve expert-level bias detection. Further, we will enhance the narrative and rely on expert-vetted content instead of BABE labels.
  \item A comprehensive evaluation of player experience, fun, and motivation, particularly concerning the tutorial and each game mode. A UX study would pinpoint and correct UX discrepancies, refining the experience to mirror a genuine GWAP. To measure motivation and enjoyment, either a qualitative UX study with open-ended questions or a quantitative UX study, such as the Single Ease Question \cite{sauroComparisonThreeOnequestion2009} to evaluate task difficulty or the Intrinsic Motivation Inventory \cite{mcauleyPsychometricPropertiesIntrinsic1989} to measure player motivation, are well-suited \cite{segundodiazBuildingBlocksCreating2022}.
  \item Launching an online version of the game to measure unpaid player interactions. This would provide insights into annotation efficiency and scalability without monetary incentives while creating an extensive dataset without further financial investment. In line with \textcite{tileAttackMentrics}, we can measure player engagement by considering lifetime judgments, average judgments per player, average lifetime play, monthly active users, retention, and throughput. Incorporating pre- and post-tutorial assessments can also determine the tutorial's impact. The impact of delayed feedback on player retention is especially interesting. Longitudinal post-gameplay studies can shed light on lasting learning effects. We further need to evaluate if quick, automatic decisions in the game modes \textit{Co-Op} and \textit{Quick Words} create stronger biases in the annotations \cite{sys1sys2InterventionsMoravec2020}.
  \item Applying a more sophisticated latent class model that integrates personal backgrounds, biases, experience, and skill levels \cite{NIPS2010_0f9cafd0, yanActiveLearningCrowds2011} to create labels. The game could, additionally to the player skill, select whose annotations to take into the final dataset based on player backgrounds to ensure a more diverse and balanced dataset, especially politically and culturally.
\end{enumerate}

Future game iterations could benefit from integrating AI-driven explanations, such as OpenAI's ChatGPT.\footnote{\url{https://openai.com/blog/chatgpt}}
Our experience shows that ChatGPT’s ability to detect biased phrases is limited.
However, it can provide explicit feedback by explaining why something may be considered biased after annotation tasks.
While models such as GPT can generate labels, we believe assessing bias through human perception will be necessary, especially by including different opinions to develop and constantly evaluate a fair AI.

Like any online community, discussion threads in News Ninja require moderation.
A report button and manual moderation of interactions are necessary, as models designed to detect hate speech might mistakenly label biased discussion content as hateful.

GWAPs, if executed proficiently, are a powerful tool for data-intensive, complex research areas if they succeed in making the process enjoyable.
They may increase participation rates, improve data quality, and provide a valuable educational experience for players.
Such methodologies can extend to various crowdsourced data collection tasks beyond linguistic bias detection.
For example, we plan to use the News Ninja system to incorporate other types of media bias, misinformation, or manipulative language with corresponding lessons, storylines, and game modes.

\section{Conclusion}
This work introduces News Ninja, the first functional Game With A Purpose (GWAP) designed to educate players on detecting linguistic bias in news texts and to gather annotations to aid in automatic bias detection.
News Ninja translates annotation guidelines into an interactive tutorial with direct feedback by applying frameworks from serious games and gamification.
We describe the design and integration of the annotation task into different game mechanics and modes.
Following a qualitative pre-test to assess player experience, a quantitative study was conducted to collect annotations via the game.
The quality of annotations is evaluated by comparing player-generated labels against those from crowdsourcers and experts.
The News Ninja dataset outperforms analogous linguistic bias datasets while achieving results comparable to experts, suggesting News Ninja as a promising approach for collecting annotations on linguistic bias.
Furthermore, News Ninja exhibits potential for scalability, adaptability, and applications in educational settings.

\begin{acks}
This work was supported by the Hanns-Seidel Foundation (\url{https:// www.hss.de/}), the German Academic Exchange Service (DAAD) (\url{https://www.daad.de/de/}), the Bavarian State Ministry for Digital Affairs in the project XR Hub (Grant A5-3822-2-16), and partially supported by JST CREST Grant JPMJCR20D3 Japan.
None of the funders played any role in the study design or publication-related decisions.
\end{acks}

\printbibliography
\appendix

\section{Demographic Survey} \label{sec:app:demo}
\begin{enumerate}
\item What gender do you identify with? (Woman, Man, Diverse, Prefer not to say)
\item What is your age? (Input field for number)
\item What is the highest level of education you have completed?
(8th grade, Some high school, High school graduate, Vocational or technical school, Some college, Associate degree, Bachelor’s degree, Graduate work, Ph.D., I prefer not to say)
\item What is the level of your English proficiency? (Proficient, Independent, Basic)
\item Do you consider yourself to be liberal, conservative, or somewhere in between? Please slide to record your response. (Very liberal to Very conservative, -10 to 10 point slider)
\item How often on average do you check the news? (Never, Very rarely, Several times per month, Several times per week, Every day, Several times per day)
\item What news outlets do you consume? (Selection through checkboxes with free text field below)
\end{enumerate}

\section{UX Study Open Questions} \label{sec:app:uxquestions}
\begin{enumerate}
\item What was your first impression when you entered the game?
\item How was your experience within the game?
\item Where did you struggle?
\end{enumerate}

\section{Game Mechanics}
\begin{longtable}{>{\raggedright}p{2.3cm}>{\raggedright}p{4.1cm}p{5.1cm}} 
\caption{Game mechanics employed by News Ninja and their purpose and potential effects, with higher engagement translating to higher amounts of collected data.} \label{tab:allmechanics} \\
\toprule
\textbf{Game Mechanic} & \textbf{News Ninja Adaptation} & \textbf{Purpose and Possible Effects} \\
\midrule
\endfirsthead

\caption[]{(continued)} \\
\toprule
\textbf{Game Mechanic} & \textbf{News Ninja Adaptation} & \textbf{Purpose and Possible Effects} \\
\midrule
\endhead

\midrule
\endfoot

\bottomrule
\endlastfoot

Direct Feedback & Color-codes outlines and highlights on word and sentence level & Enable players to learn from their input and repetition; Increase bias detection skills and IAA \\
Delayed feedback and uncertainty & Yellow outlines and highlights on word and sentence level; Notification and higher reward on formation of ground truth & Inform players that ground truth hasn't formed yet; Increase motivation to return \\
Guidance through pedagogical agent & Plant guiding through the tutorial and speaking motivating to players during gameplay & Transmit learning objectives; Increase bias detection skills, IAA, and motivation \\
Narrative & Plant tells story of player as intern in a news outlet & Increase enjoyment, motivation and learning effects through context related to the learning objectives; Increase understanding of media bias and engagement\\
Tutorial & Slowly increasing complexity through levels and new mechanics & Sense of progression and achievement; Increase bias detection skills and IAA\\
Appeal to higher meaning & Stating the purpose of News Ninja and effect in the world & Increase intrinsic altruistic motivation and engagement\\
Rewards and penalties & Experience points, in-game currency, rise in skill level, time penalties & Increase fun, extrinsic motivation, learning effects, and engagement\\
Ownership: Assessment & Skill level bar, feedback after game round, feedback after annotation & Show players their capabilities and increase of them over time; Increase intrinsic motivation and engagement \\
Progression & Level, Skill level bar, unlocking content and game modes & Increase fun and intrinsic and extrinsic motivation; Increase engagement and bias detection skills\\
Collaboration and Competition & Game mode \textit{Co-Op}; Group Mission & Increase fun and engagement through social collaboration \\
Time pressure & Game mode \textit{Quick Words} & Increase fun and annotation collection on word level\\
Ownership: Collecting & Experience points, in-game currency, rise in skill level & Increase fun, extrinsic motivation, and engagement\\
Responsibility & Stating the game's mission, especially through the group mission & Increase intrinsic motivation and engagement\\
Discussion & Discussing bias annotations of sentences with other players & Reflection and motivation through social interactions; Increase fun and intrinsic motivation, potentially increasing bias detection skills\\
\end{longtable}

\end{document}